\documentclass[12pt]{article}
\usepackage{url} 

\usepackage{latexsym, amssymb, amscd, amsthm, amsxtra, amsmath,amsthm, mathtools, mathrsfs}
\usepackage{graphics, graphicx, color}
\usepackage{natbib}
\usepackage{ifpdf}
\usepackage[format=hang,indention=-1cm,small]{caption}
\usepackage{subcaption}
\usepackage{multirow}
\usepackage{kotex}
\usepackage{hyperref}
\usepackage{epsfig}
\usepackage[linesnumbered,ruled,vlined]{algorithm2e}
\usepackage{arydshln}
\usepackage{ulem}
\usepackage{booktabs}
\usepackage{bm}
\usepackage{xcolor}
\usepackage{xparse}
\usepackage{chngcntr}

\numberwithin{equation}{section} 

\newcommand{\blind}{0}

\newtheorem{thm}{Theorem}[section]
\newtheorem{cor}[thm]{Corollary}
\newtheorem{lem}[thm]{Lemma}
\newtheorem{prop}[thm]{Proposition}

\theoremstyle{definition}
\newtheorem{defn}{Definition}[section]
\newtheorem*{defn*}{Definition}

\theoremstyle{remark}
\newtheorem{remark}{Remark}[section]
\newtheorem*{remark*}{Remark}

\newcommand{\Mc}{\mathcal{M}}

\newcommand{\Oc}{\mathcal{O}}
\newcommand{\Pc}{\mathcal{P}}

\newcommand{\Vc}{\mathcal{V}}

\newcommand{\Xc}{\mathcal{X}}
\newcommand{\Yc}{\mathcal{Y}}

\def \Eb {\mathbb{E}}
\def \Rb {\mathbb{R}}

\def \Pb {\mathbb{P}}
\def \Sb {\mathbb{S}}

\def\vv{\mathbf v}

\def\Vv{\mathbf V}

\def\0v{\mathbf{0}}

\def\R{\mathbb{R}}

\def\1v{\mathbf 1}
\def\0v{\mathbf 0}

\DeclareMathOperator{\er}{er}

\DeclareMathOperator{\diag}{diag}
\DeclareMathOperator{\tr}{tr}

\DeclareMathOperator{\vecd}{vecd}
\DeclareMathOperator{\cov}{Cov}
\DeclareMathOperator{\var}{var}
\DeclareMathOperator{\col}{col}

\DeclareMathOperator*{\argmin}{arg\,min}

\begin{document}

\def\spacingset#1{\renewcommand{\baselinestretch}%
{#1}\small\normalsize} \spacingset{1}


\if0\blind
{
  \title{\bf Robust and Differentially Private Principal Component Analysis}
  \author{{Minwoo Kim}\\
    Department of Statistics, Seoul National University\\
	and \\
	Sungkyu Jung\thanks{
        This work was supported by the National Research Foundation of Korea (NRF) grants funded by the Korea government (MSIT) (No. 
        RS-2023-00218231,  
        RS-2023-00301976,  
        RS-2024-00333399). 
	}
    \hspace{.2cm}\\
	Department of Statistics and Institute for Data Innovation in Science, \\Seoul National University}
  \date{October 12, 2025}
  \maketitle
} \fi

\if1\blind
{
  \bigskip
  \bigskip
  \bigskip
  \begin{center}
    {\LARGE\bf Title}
  \end{center}
  \medskip
} \fi

\bigskip
\begin{abstract}
Recent advances have sparked significant interest in the development of privacy-preserving Principal Component Analysis (PCA). However, many existing approaches rely on restrictive assumptions, such as assuming sub-Gaussian data or being vulnerable to data contamination. Additionally, some methods are computationally expensive or depend on unknown model parameters that must be estimated, limiting their accessibility for data analysts seeking privacy-preserving PCA. In this paper, we propose a differentially private PCA method applicable to heavy-tailed and potentially contaminated data. Our approach leverages the property that the covariance matrix of properly rescaled data preserves eigenvectors and their order under elliptical distributions, which include Gaussian and heavy-tailed distributions. By applying a bounded transformation, we enable straightforward computation of principal components in a differentially private manner. Additionally, boundedness guarantees robustness against data contamination. We conduct both theoretical analysis and empirical evaluations of the proposed method, focusing on its ability to recover the subspace spanned by the leading principal components. Extensive numerical experiments demonstrate that our method consistently outperforms existing approaches in terms of statistical utility, particularly in non-Gaussian or contaminated data settings.
\end{abstract}

\noindent%

\newpage
\spacingset{1.5} 


\section{Introduction}
As the amount and variety of data being collected grow, understanding the underlying structure of the data becomes increasingly important.
Principal component analysis (PCA) is a fundamental yet powerful method for dimensionality reduction, and identifies the principal component directions that define a low-dimensional subspace capturing the maximum-variance inherent in the data. 


In recent decades, limiting the disclosure of sensitive personal information in data analysis has become an important issue.
To address this, \textit{differential privacy} (DP) has emerged as the de facto standard for ensuring privacy guarantees \citep{Cummings2024Advancing, su2025statDP}.
The original notion of DP, called $\varepsilon$-DP, was first introduced by \cite{dwork2006calibrating}, and is defined as follows.
Consider a randomized map (or mechanism) $ \Mc:\Xc^n \to \Yc $ that outputs a randomly perturbed statistic from a dataset. 
The randomized mechanism $\Mc$ is said to be $ \varepsilon $-DP for some $\varepsilon > 0$ if for any neighboring datasets $S = (x_1, \dots, x_{n-1}, x_n) \in \Xc^n$ and $S' = (x_1, \dots, x_{n-1}, x_n') \in \Xc^n$, which differ in exactly one record, it holds that  
\[
\log \left(
\frac{\Pb(\Mc(S) \in E)}{\Pb(\Mc(S') \in E)}
\right) \le \varepsilon,
\quad \text{for any measurable $E \subset \Yc$.}
\]
The privacy parameter $\varepsilon \in (0, \infty)$ represents the extent to which the output distributions remain indistinguishable when a single individual's record is  replaced arbitrarily.
Several extensions of DP beyond $\varepsilon$-DP have been proposed to provide more flexible frameworks for measuring indistinguishability \citep{bun2016concentrated,mironiv2017renyi,dong2022gdp}. 
To design a mechanism satisfying a given level of DP, a common approach is to add properly scaled random noise to the statistic of interest. 
For example, a real-valued randomized mechanism $\Mc(S) = \sum_{i=1}^n x_i + \zeta$ satisfies  $\varepsilon$-DP if $\zeta$ follows a centered Laplace distribution with scale parameter $ \sup_{x \in \Xc}|x| / \varepsilon $ \citep{dwork2014algorithmic}.
In addition, various DP mechanisms designed to enhance utility have been proposed,  including the propose-test-release \citep{dwork2009ptr}, the exponential mechanism \citep{mcsherry2007mechanism}, the smooth-sensitivity mechanism \citep{nissim2007smooth}, and the private stochastic gradient descent algorithm \citep{abadi2016DLDP}. Building on these tools, researchers have recently developed a wide range of DP methods, including histogram construction, contingency table analysis, regression, and hypothesis testing \citep{kim2025differentially,wang2018revisiting,yu2014differentially,jung2024highly,awan2025differentially}. A brief review on DP is provided in Section \ref{sec:DP}.

Alongside privacy, the importance of \textit{robustness} has been increasingly emphasized.
In practice, data may be heavy-tailed, deviate from Gaussian assumptions, or contain outliers and anomalies. 
In addition to such natural irregularities, a malicious user can compromise  model training by injecting poisoned data into the training set \citep{xiao2015feature, chen2017targeted}.  
Thus, developing statistical methods that are both private and robust has become crucial for modern data analysis.
However, DP alone does not guarantee robustness, as differentially private models can remain vulnerable to adversarial attacks \citep{liu2021robust}. Moreover, without accounting for robustness, the presence of outliers can substantially inflate the noise scale required by DP methods (e.g., $\sup_{x\in\Xc} |x|/\varepsilon$), leading to a severe loss of utility. 

In this paper, we develop a \textit{robust and differentially private PCA}. 
The proposed method is simple, easily computable, and allows an arbitrary number of PC directions. Robustness of the proposed PCA is ensured by utilizing two recently developed tools: the generalized spatial sign of \cite{raymaekers2019generalized}  and the matrix-variate extension of Kendall's tau studied in \cite{han2018eca}.
By adapting the resulting robust PCA framework to satisfy DP, our method simultaneously achieves both privacy and robustness.

A generalized spatial sign $g: \Rb^d \to \Rb^d$ has the form 
$g(t) = \xi(\|t\|_2) \cdot t / \|t\|_2 $ for some positive-valued function $\xi$.
Thus, $ g(t) $ can be viewed as a rescaling of the given vector $t$ while maintaining its 
direction.
%
%
\cite{raymaekers2019generalized} studied a generalized covariance matrix defined as $\Sigma_g = \Eb[g(X-\mu)g(X-\mu)^\top]$, where $\mu = \Eb(X)$. 
When $X$ follows an elliptical distribution, which includes Gaussian and heavier-tailed distributions, the eigenvectors of $\Sigma_g$ coincide with those of the scatter-matrix $\Sigma$ of $X$, thus PC directions can also be  obtained from $\Sigma_g$. 
However, under the DP setting, privately estimating $ \mu $ introduces additional noise, which leads to inefficiency and degrades the performance of the final PCs.
To overcome this issue, we adopt an idea of multivariate Kendall's tau  defined as $ K = \Eb[(X-\widetilde X)(X-\widetilde X)^\top / \|X - \widetilde X\|_2^2] $, where $ \widetilde X $ is an independent copy of $X$ \citep{han2018eca}.
Considering the difference between independent copies circumvents the issue of estimating $ \mu $.
We thus extend $ K $ by replacing the normalization with the generalized spatial sign function $g$, defining the generalized multivariate Kendall's tau matrix as 
$K_g = \Eb[g(X - \widetilde X)g(X - \widetilde X)^\top]$. 
As with $ \Sigma_g $, $ K_g $ also shares the eigenvectors with $\Sigma$.
This justifies changing our focus to the eigenvectors of $K_g$ for the purpose of PCA.
For a non-private sample version, we consider a second order $U$-statistic $ \widehat K_g$ as an estimator of $ K_g $. 
To achieve differential privacy, a properly scaled Gaussian noise is added to each element of $ \widehat K_g $ resulting in a perturbed matrix $\widetilde K_g$.
Finally, private PCs can then be obtained by extracting eigenvectors from $ \widetilde K_g $.

In our proposal, we restrict the class of $ g $ to only contain those with a bounded $\sup$-norm to make the variance of the additive Gaussian noise finite.
Also, with bounded $ g $, the proposed private PCA is endowed with the robustness against heavy-tailed distribution and data contamination.

Under an elliptical data model and bounded $g$, we show that the proposed private PCA method consistently estimates the population eigen-subspace. 
We also theoretically verify its robustness against arbitrary data contamination. 
Our numerical studies provide evidence that proposed PCA methods behave well under non-Gaussian and corrupted data settings.
In these cases, the proposed method outperforms the other competing methods in terms of eigensubspace recovery in simulations.

To the best of our knowledge, few existing works for PCA address both robustness and privacy. 
From the perspective of the optimization problem, \cite{maunu2022stochastic} privately solved the $L_1$-norm PCA problem by the geodesic gradient descent on the space of orthogonal matrices under the inlier-outlier model.
However, they only consider deterministic data models, and none of the statistical utilities, such as robustness to data corruption or error bound on subspace recovery, are provided.
On the other hand, some authors have converted robust statistical estimators into  private versions.
\cite{liu2022robustdp} generalized the propose-test-release \citep{dwork2009ptr} and developed a general private algorithm covering various statistical problems, including robust PCA.
\cite{asi2023robustness} proposed a private PCA by converting a robust estimator into 
a private one by utilizing the inverse sensitivity mechanism of \cite{asi2020instance}.
Under Gaussian assumption, both \cite{liu2022robustdp} and \cite{asi2023robustness} provide estimators for the first PC direction that achieve a minimax lower bound of expected risk up to logarithmic factors.
However, both algorithms are computationally ineffective, which is difficult to implement in practice, and it is not straightforward to extend these methods to extract more than one PC direction.

The remainder of the paper is structured as follows.
Section \ref{sec:DP} reviews the basic notions of differential privacy, and commonly used notations are introduced in Section \ref{sec-notation}.
In Section \ref{sec:methods}, we propose a private PCA method based on the generalized Kendall's tau matrix.
Section \ref{sec-3} provides theoretical guarantees, including error bounds and robustness.
In particular, we study two specific choices of $g$; spherical transformation and winsorization, in more details.
Section \ref{sec:numerical-study} presents numerical comparisons among the proposed methods and existing methods by reporting the average loss of eigensubspace estimation under various settings. 
Additionally, we provide a visualization of the Europe genetic data onto the first two private principal components.
All proofs, technical details and additional numerical studies are deferred to the appendix.

\subsection{Preliminaries on differential privacy}\label{sec:DP}

Let $S = (x_1, \dots, x_n)  \in \Xc^n$ be a dataset collected from a data space $\Xc \subset \Rb^{d}$. 
For any two datasets $S = (x_1, \cdots, x_n)$ and $S' = (x_1', \cdots, x_n')$ of equal size, we say $ S $ 
and $ S' $ are neighbors if they differ by a single record at most, i.e., 
if there exists an $i \in \{1,\ldots,n\}$ such that $x_i \neq x_i'$ and $x_j = x_j'$ for all $j \neq i$.
We denote this neighboring relation as $S \sim S'$.

Let $ \theta: \Xc^n \to \Rb^d $ be a map taking a dataset $ S = (x_1, \dots, x_n) $ as an input, 
and outputs $ \theta(S) $, the statistic of interest.
In differential privacy, a randomly perturbed version of $\theta(S)$ is released 
to the user to prevent the disclosure of sensitive information in the dataset $S$.
This randomly perturbed version of $\theta(\cdot)$ is often called a randomized mechanism. 
If $\Mc: \Xc^n \to \Rb^d$ is a randomized mechanism, then its output $\Mc(S)$ is an $\Rb^d$-valued random vector.
The framework of differential privacy provides a rigorously quantifiable measure of privacy protection of such a mechanism $ \Mc $. 

\begin{defn}[\cite{dwork2006calibrating}]
For $ \varepsilon > 0 $ and $ \delta \in [0, 1) $, a randomized mechanism $\Mc: \Xc^n \to \Rb^d$ satisfies $(\varepsilon, \delta)$-\textit{differential privacy} (DP) if for any neighboring datasets $S \sim S'$, and for any event $E \subset \Rb^d$, 
\[
    \Pb(\Mc(S) \in E) \le e^\varepsilon  \Pb(\Mc(S') \in E) + \delta.
\]
Here, the randomness only depends on the mechanism $\Mc$.
If $\delta = 0$, $\Mc$ satisfies $\varepsilon$-DP. 
\end{defn}

The parameters $ (\varepsilon, \delta) $ are called privacy parameters, or privacy budget.
Smaller values of $ \varepsilon$ and $ \delta $ make it more difficult to distinguish between the outputs of $ \Mc(S) $ and $ \Mc(S') $, thereby making it harder for attackers to infer sensitive information in the data from the released output.

Next, we introduce a noise-additive mechanism called Gaussian mechanism. 
Let $\theta: \Xc^n \to \Rb^d$ be a statistic of interest.
Then, the Gaussian mechanism with noise scale $\sigma > 0$ is defined as $ \Mc_G(S; \sigma) = \theta(S) + N_d(0, \sigma^2I_d) .$
A higher value of $ \sigma $ provides stronger privacy protection, while a lower value of $ \sigma $ results in weaker privacy. 
To determine the minimum level of noise necessary to ensure $(\varepsilon, \delta)$-DP, we need a notion of sensitivity of $ \theta $.
The $\ell_2$-sensitivity of $\theta$ is defined as $\Delta_{2}(\theta) = \sup_{S \sim S'} \|\theta(S) - \theta(S')\|_2,$ where the supremum is taken over all pairs of neighboring datasets of $\Xc^n$. 
Thus, the sensitivity measures the maximal perturbation of $ \theta $ after an arbitrary change of one data point.
To make $ \Mc_G(S; \sigma) $ satisfy $ (\varepsilon, \delta) $-DP, $ \sigma $ should be calibrated proportional to $ \Delta_2(\theta) $, as stated next.

\begin{prop}[\cite{dwork2014algorithmic}]\label{prop:gaussian-mech}
    Let $\Mc_G(S; \sigma) = \theta(S) + N_d(0, \sigma^2 I_d)$ be an additive Gaussian mechanism for $\theta: \Xc^n \to \Rb^d$ with $\sigma > 0$. 
    Let $\varepsilon> 0$ and $ \delta \in (0, 1)$ be privacy parameters.
    If $\sigma \ge \frac{\Delta_2(\theta)\sqrt{2\ln(1.25/\delta)}}{\varepsilon}$, then $\Mc$ is $(\varepsilon, \delta)$-DP.
\end{prop}

We close this section by presenting one prominent property of DP called the \textit{post-processing} property. 
It says that a mechanism's privacy guarantee is not affected by data-independent manipulation of output once it has been privately released.
Suppose that a mechanism $ \Mc$ satisfies $ (\varepsilon, \delta) $-DP, 
then for any (possibly randomized) map $ T$ that takes output of $ \Mc $ as an input, the post-processed map $ T \circ\Mc$ also satisfies $ (\varepsilon, \delta) $-DP.

\subsection{Notation} \label{sec-notation}
We introduce notations frequently used in the paper.
We denote  $\mbox{Sym}(d) $ for a set of $ d \times d $ real-valued symmetric matrices.
For a vectorization of a symmetric matrix, we use the function
$\vecd: \mbox{Sym}(d) \to \mathbb{R}^{d(d+1)/2}$, which maps 
$Y \mapsto (\diag(Y)^{\top}, \sqrt{2} \mbox{offdiag}(Y)^{\top})^{\top}$  \citep{schwartzman2016lognormal}. 
The inverse function $ \vecd^{-1}: \Rb^{d(d+1)/2} \to \mbox{Sym}(d) $ is also well-defined.
We denote the set of matrices consisting of $m$-orthonormal vectors as 
$\Oc(d, m) = \{U \in \Rb^{d \times m}:U^\top U = I_m\}$ and simply write $\Oc(d) \equiv \Oc(d, d)$ 
for the set of orthonormal matrices.
For a matrix $A$, $\col(A)$ denotes the column space of $A$.
For $ a, b \in \Rb $, we denote $a \lesssim b$ if there exists an absolute constant $ C > 0 $ such that $ a < Cb $. 
We denote $a \asymp b$ if $ a \lesssim b $ and $ b \lesssim a $.

\section{Proposed methods} \label{sec:methods}

\subsection{Generalized multivariate Kendall's tau matrix} 

Consider a pair of univariate random variables $(x, y)$ and let $(\tilde x, \tilde y)$ be its independent copy.
The population (univariate) Kendall's tau, or Kendall's rank correlation between $x$ and $y$ is defined as 
\[
\tau
= \cov(\mbox{sign}(x-\tilde x), \mbox{sign}(y - \tilde y))
= \Eb \left[\mbox{sign}(x-\tilde x)\mbox{sign}(y - \tilde y)\right], 
\]
where $\mbox{sign}(y) = y / |y|$.
A multivariate extension of the univariate sign is the spatial sign, which maps $Y \in \Rb^d$ to its normalized direction $ Y / \|Y\|_2 $.
This extension leads to the multivariate version of Kendall's tau.
Let $ X \in \Rb^d $ be a random vector of the observation and $ \widetilde X $ be the independent copy of $ X $. 
The population multivariate Kendall's tau matrix is defined as
\begin{equation} 
    K 
    = \Eb \left[\left(\frac{X - \widetilde X}{\|X - \widetilde X\|_2}\right)\left(\frac{X - \widetilde X}{\|X - \widetilde X\|_2}\right)^\top \right].
\end{equation}
Importantly, eigenvectors of $K$ are the same as those of $\cov(X)$ when $X$ follows an elliptical distribution \citep{visuri2000sign, han2018eca}.
The spatial sign is a self-normalization process that  mitigates the effect of gross outliers in the observed data.
Thus, the eigenvectors of an estimator $\widehat K$ of $K$ are valid and robust PC direction estimators.

Recently, \cite{raymaekers2019generalized} have further extended the spatial sign to a general class of transformations.
For a positive valued scale function $ \xi: (0, \infty) \to (0, \infty) $, consider a map $ g_\xi: \R^d \to \R^d $ defined as 
\begin{equation} \label{eq:g}
    g_\xi(t) = \xi(\|t\|_2)\cdot \frac{t}{\|t\|_2}.
\end{equation}
We call $g_\xi$ generalized spatial sign.
Clearly, the spatial sign is a special case of the generalized spatial sign with $\xi \equiv 1$.
In general, $ g_\xi $ transforms a given point $ t \in \R^d $ by rescaling through $ \xi $, so that $g_\xi(t)$ preserves the direction of $t$, but its length becomes $\xi(\|t\|_2)$.
We propose a natural extension of multivariate Kendall's tau by replacing the spatial sign with a generalized spatial sign as follows.

\begin{defn} 
    Let $ g_\xi: \R^d \to \R^d $ be the transformation defined in \eqref{eq:g} with a scale function 
    $ \xi: (0, \infty) \to (0, \infty) $.
    The \textit{generalized multivariate Kendall's tau} matrix with respect to $ g_\xi$ is defined as
    \[
        K_{g_\xi} = \Eb_{X, \widetilde X}\left[ g_\xi\left( \frac{X - \widetilde X}{\sqrt{2}}\right) 
        g_\xi\left( \frac{X - \widetilde X}{\sqrt{2}}\right)^\top ~ \right].
    \]
\end{defn}

The eigenvectors of $K_{g_\xi}$ and $\cov(X)$ are the same if $X$ follows an elliptical distribution (cf. Section \ref{sec:robustPCA-elliptical}) as long as $g_\xi$ has the form of \eqref{eq:g}.
Here, we give some examples of $g_\xi$.
First, if we set $ \xi(s) = s $, then $ g_\xi(t) = t $. In this case, $ K_{g_\xi} = \cov(X) $, the usual covariance matrix. 
Next example is the spatial sign
\begin{equation} \label{eq-gsph}
    g_{sph}(t) := \frac{t}{\|t\|_2},
\end{equation}
which corresponds to $\xi(s) = 1$.
We use subscript \textit{sph} to give an emphasis on the fact that the spatial sign is a projection map onto the unit sphere $ \Sb^{d-1} $, and we call $g_{sph}$ spherical transformation.
Another important example is winsorization.
For a given radius $ r > 0 $, let $ \xi(s) = \min(r, s) $.
Then, its corresponding transformation, denoted as $ g_{wins}^{(r)} $, is defined as 
\begin{equation} \label{eq-gwins}
 g_{wins}^{(r)} (t)  = \min(r, \|t\|_2) \cdot \frac{t}{\|t\|_2} 
    =  \begin{cases}
        ~t & \text{if $\|t\|_2 \le r$},  \\
        ~r \cdot t/\|t\|_2  & \text{if $\|t\|_2 > r$}.
    \end{cases}
\end{equation}
That is, the winsorization map leaves the point $t$ unchanged if $t$ lies within the radius-$r$ ball. 
Otherwise, it projects $ t $ onto the boundary of the radius-$r$ ball.
For more examples of generalized spatial signs $ g_\xi $, see \cite{raymaekers2019generalized} and \cite{leyder2024generalized}.  
In the rest of the paper, we use $g$ as the generic notation for $g_\xi$, representing a generalized spatial sign. 

We say a generalized spatial sign $ g $ is \textit{bounded} if
\[
    \|g\|_\infty := \sup \{\|g(t)\|_2: t \in \Rb^d\} < \infty.
\]
We restrict our attention to bounded generalized spatial signs. 
In particular, 
the spherical transformation $g_{sph}$ is bounded since $ \|g_{sph}\|_{\infty} = 1 $, and the the winsorization map is also bounded with $ \|g_{wins}^{(r)}\|_\infty = r$. 
In contrast, the identity map $g(t) = t$, yielding the usual covariance matrix, is unbounded. 

Letting $S=(X_1, \dots, X_n)$ be a simple random sample, we use the second order $ U $-statistics as an estimator of $ K_g $:
\begin{equation} 
    \widehat K_g(S) = \frac{2}{n(n-1)} \sum_{i < j} g\left(\frac{X_j - X_i}{\sqrt{2}}\right)
    g\left(\frac{X_j - X_i}{\sqrt{2}}\right)^\top.
\end{equation}
The boundedness of $ g $ makes  $ \widehat K_g $ robust to data contamination (see Theorems \ref{thm:bp-Kendalltau} and \ref{thm:corrupted} in Section \ref{sec-theory-robust}).
Furthermore, bounded $g$ also implies that $ \widehat K_g $ has finite sensitivity. 
As a consequence, adding a Gaussian noise  to $ \widehat K_g $ provides a differentially private estimator of $ K_g $, where the standard deviation of additive Gaussian noise is scaled in proportion to the sensitivity of $ \widehat K_g $. 
A detailed procedure is given in the next subsection.

\subsection{Proposed differentially private PCA} \label{sec:private-pca}

We propose a differentially private PCA mechanism based on $ \widehat K_g $.
Denote the sensitivity of $ \widehat K_g $ with respect to the Frobenius norm by $ \Delta_F(\widehat K_g) = \sup_{S \sim S'} \|\widehat K_g(S) - \widehat K_g(S')\|_F $.
If $g$ is bounded, then the sensitivity is bounded and satisfies $ \Delta_F(\widehat K_g) \le 4\|g\|_\infty^2/n $, even though the domain for data $S$ is unbounded; see Appendix \ref{pf-prop2.1}. 
Although $ \widehat K_g $ is a matrix-valued estimator, the vectorization operator enables us to apply the standard vector-valued Gaussian mechanism of Proposition \ref{prop:gaussian-mech}.
For a dataset $S = (x_1, \dots, x_n) \in (\Rb^d)^n$ and $ \sigma >0 $, we define a randomized mechanism 
\[
\widetilde K_g(S; \sigma) :=   
\frac{2}{n(n-1)} \sum_{i < j} g\left(\frac{x_j-x_i}{\sqrt{2}}\right)g\left(\frac{x_j-x_i}{\sqrt{2}}\right)^\top + \vecd^{-1}(\xi),
\]
where $\xi \sim N_{d(d+1)/2}(0, \sigma^2 I_{d(d+1)/2}). $
We then define $ \widetilde V_g(S; \sigma) \in \Oc(d) $ as the eigenvector matrix of $\widetilde K_g(S; \sigma)$, whose $i$th column is the eigenvector associated with the $i$th largest eigenvalue.
Let $ \sigma_{\varepsilon, \delta} := \frac{4\|g\|_{\infty}^2 \sqrt{2 \ln(1.25/\delta)}}{n\varepsilon} $.

\begin{prop} \label{prop-privKg}
    For any $d \ge 2$ and sample size $n \in \mathbb N$, let the domain for the data set $ S $ be $(\Rb^d)^n $.
    For all $ \sigma \ge \sigma_{\varepsilon, \delta} $, 
    $\widetilde K_g(S; \sigma)$ satisfies $ (\varepsilon, \delta) $-DP.    
    Furthermore, $ \widetilde V_g(S; \sigma) $ also satisfies $ (\varepsilon, \delta) $-DP.
\end{prop}

In Proposition \ref{prop-privKg}, the privacy guarantee of $ \widetilde{K}_g $ is established by the additive Gaussian mechanism, while $ \widetilde{V}_g $ inherits DP guarantee through the post-processing property.
Since a smaller value of $ \sigma $ is preferable under the same privacy budget, we fix $\sigma = \sigma_{\varepsilon, \delta}$ in the rest of the paper. 
The matrix of private principal component direction estimators $\widetilde V_{g}(S) \equiv \widetilde V_{g}(S; \sigma_{\varepsilon, \delta})$ is called  \textit{$ g $-DPPCA}.
In Sections \ref{sec-3} and \ref{sec:numerical-study}, we investigate the statistical efficiency of $\widetilde V_g(S)$ in terms of eigensubspace recovery and robustness against contamination, from both theoretical and numerical perspectives.

\begin{remark}
    To construct a private estimator for the eigenvalue of the population covariance $ \Sigma $, the idea used in non-private robust PCA methods \citep{leyder2024generalized} can be employed.
    As an illustrative example, consider obtaining the largest eigenvalue. First, privately estimate the first eigenvector by using the $ g $-DPPCA, say $ \tilde v_1 $. 
    Then, for a given observation $ (x_1, \dots, x_n) $, a private estimator of the first eigenvalue can be obtained by privately estimating
    the variance of the projected score $ \{\tilde v_1^\top x_i\}_{i=1}^n $.
    The rest of the eigenvalues can be obtained by similar process. 
\end{remark}

\begin{figure}[h]
    \centering
    \begin{subfigure}{0.48\textwidth}
        \centering
        \includegraphics[width=\linewidth]{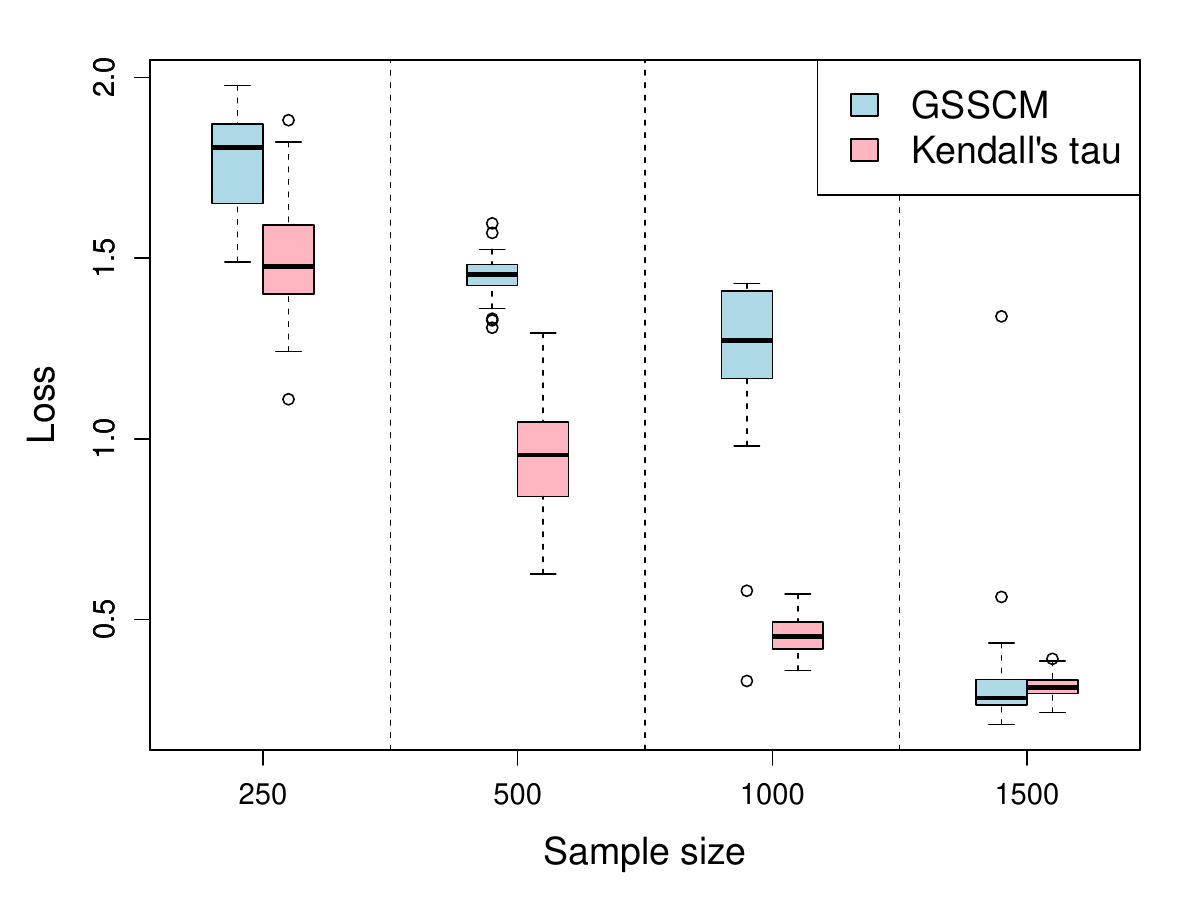}
        \caption{Comparison to $\widehat \Sigma_g$}
        \label{fig:GSSCM-PMWM}
    \end{subfigure}
          \hfill
    \begin{subfigure}{0.48\textwidth}
        \centering
        \includegraphics[width=\linewidth]{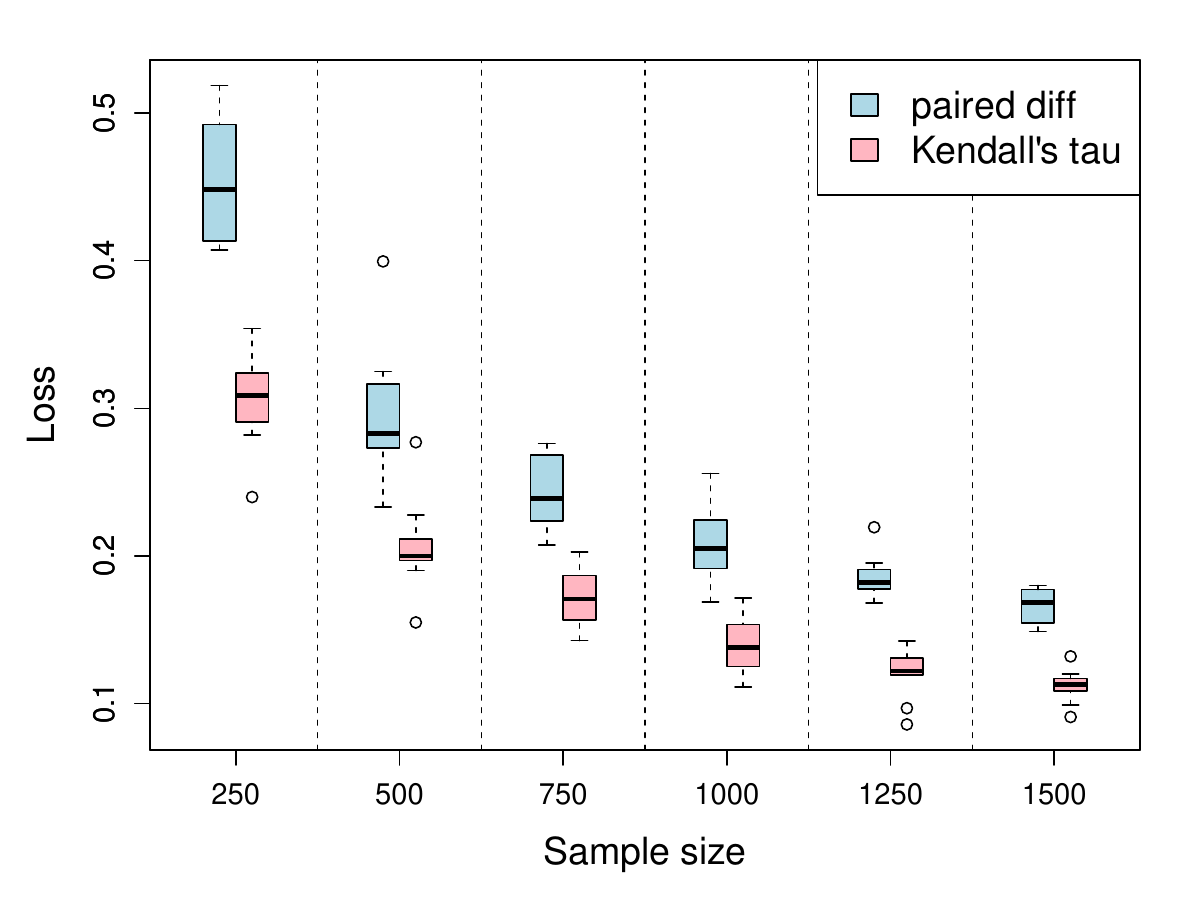}
        \caption{Compariosn to $ \breve{K}_g $}
        \label{fig:paired_sph}
    \end{subfigure}
    \caption{
        Datasets were sampled from a 25-dimensional Gaussian distribution, specified in Section \ref{sec:sim-study}. We report the loss $ \|\widehat V_2 \widehat V_2^\top - V_2 V_2^\top\|_F $, where $ \widehat V_2$ collects the first two estimated PCs, and $V_2$ is the population counterpart. 
        (a) Comparison between private PCA based on GSSCM with a private location estimator and the proposed $g$-DPPCA. (b) Comparison of non-private PCA using $\widehat K_g$ and $\breve K_g$.
In both cases, Kendall's tau shows superior performance.
    }
    \label{fig:label-paired_kendall}
\end{figure}

\begin{remark} \label{remark_2-2}
It is natural to consider other robust candidates  that utilize the generalized spatial sign $g$. We argue that our choice of the generalized Kendall's tau and its estimator $\widehat{K}_g(S)$ is superior  compared with the following alternatives. 
The first alternative is simply use a moment estimator for the generalized spatial sign covariance (GSSCM) matrix, defined as $\Sigma_{g} = \Eb\left[ g(X - \mu)g(X - \mu)^\top \right]$, where $\mu$ is a location parameter \citep{raymaekers2019generalized}. 
With a robust location estimator $\hat\mu$, eigenvectors of $ \widehat \Sigma_g := n^{-1}\sum_{i=1}^{n} g(X_i - \hat \mu)g(X_i - \hat \mu)^\top $ serve as robust estimators of PCs. 
However, to privatize $ \widehat \Sigma_g $, an additional privacy budget is required to privately estimate $\hat\mu$, often leading to excessive noise, as depicted in Figure \ref{fig:GSSCM-PMWM}. 
The second alternative is to directly estimate $K_g$ using paired differences: $\breve{K}_g := \frac{1}{\lfloor n/2 \rfloor} \sum_{i=1}^{\lfloor n/2 \rfloor} g(W_i) g(W_i)^\top$, where $ W_i = (X_{i + {\lfloor n/2 \rfloor}} - X_{i}) / \sqrt{2} $. However, our $U$-statistic $\widehat K_g$ shows considerably higher efficiency than $\breve{K}_g$ which effectively halves the sample size; see Figure \ref{fig:paired_sph}. 
 We therefore omit further discussion of $\widehat \Sigma_g$ and $\breve{K}_g$.

\end{remark}

\section{Statistical efficiency and robustness} \label{sec-3}

In this section, we provide theoretical results for the error bound on subspace recovery of the proposed private PCA under a broad class of distributions, including heavy-tailed ones. 

We also show that robustness is guaranteed under arbitrary data corruption.
In the final subsection, we further investigate two specific cases: spherical transformation and winsorization.

\subsection{Model} \label{sec:robustPCA-elliptical}

We give a brief overview of the elliptical distribution and then present how 
$ \widehat K_g $ enables a valid PCA procedure under the elliptical distribution.

A $d$-dimensional random vector $Z$ follows a spherically symmetric distribution if and only if $QZ \overset{d}{=} Z$ for any $Q \in \Oc(d)$, meaning that $ Z $ is 
distributionally invariant to orthogonal transformations.
For a spherically symmetric $Z$, its density is of the form of 
$ \omega(\|t\|_2) $ for some function $ \omega: [0, \infty) \to [0, \infty) $.
Consider a symmetric positive-definite matrix $\Sigma$, representing the dispersion of the distribution.
For a $d$-dimensional random vector $X$, we say $X$ follows a (continuous) \textit{elliptical distribution} with location parameter $\mu \in \Rb^d$ and dispersion matrix $\Sigma$ if $X \overset{d}{=} \mu + \Sigma^{1/2}Z$, where $Z$ follows a spherically symmetric distribution satisfying $\Pr(Z=0)=0$. 
We denote it as $ X \sim \mbox{EC}_d(\mu, \Sigma, \omega) $, where $\omega(\|\cdot\|_2)$ is the density function of $Z$.
For a detailed introduction of elliptical distributions, we refer the monograph such as \cite{fang2018symmetric}.

We consider a PCA problem 
that estimates a first few eigenvectors of dispersion matrix $ \Sigma $
from the random sample $ X_1, \dots, X_n \overset{i.i.d.}{\sim} \mbox{EC}_d(\mu, \Sigma, \omega) $.
We note that this problem is well-defined even if the moment of 
$ X $ does not exist.
When $ \cov(X) $ exists, then $ \cov(X) \propto \Sigma $ up to constant, and thus
eigenvectors of $ \Sigma $ are the same as those of $ \cov(X) $.
Under this elliptical data model, we can conduct PCA via estimating $K_g$ instead of $\Sigma$.

\begin{prop}\label{prop:gkendall}
    Let $ X \sim \mbox{EC}_d(\mu, \Sigma, \omega) $. 
    Denote the eigendecomposition of $\Sigma$ as $\Sigma = U \Lambda U^\top$, 
    where $\Lambda = \diag(\lambda_1, \dots, \lambda_d)$ and $ \lambda_1 \ge \dots \ge \lambda_d > 0 $.
    Let $g$ be a bounded generalized spatial sign. Then,
    \begin{enumerate}
        \item $K_g$ has the same eigenvectors as $\Sigma$.
        \item Suppose further that $\omega$ is a decreasing function.
        Then the order and multiplicity of eigenvectors of $K_g$ are the same as $\Sigma$.
        More precisely, we can write the eigendecomposition of $K_g$ as $K_g = U \Phi_g U^\top$
        for a diagonal matrix $\Phi_g = \diag(\phi_{g, 1}, \dots, \phi_{g, d})$, satisfying 
        $ \phi_{g, 1} \ge \dots \ge \phi_{g, d} > 0 $ and $\phi_{g, j} = \phi_{g, j+1}$ if and only if $ \lambda_{j} = \lambda_{j+1} $. 
    \end{enumerate}
\end{prop}

Note that $ g $ is orthogonally equivariant in the sense that $ g(Qt) = Q g(t)  $ for any orthogonal matrix $ Q \in \Oc(d) $.
The orthogonal equivariance of $g$ preserves eigenvectors of $\Sigma$, and the decreasing property of $\omega$ ensures the order of eigenvalues are maintained.
Similar results have been observed for some special cases of the multivariate Kendall's tau matrix \citep{marden1999some, croux2002sign, taskinen2012robustifying}.

In the light of Proposition \ref{prop:gkendall}, we shall assume 
$X \sim \mbox{EC}_d(\mu, \Sigma, \omega)$ with decreasing $\omega$ throughout. 
In addition, we denote $ \lambda_1 \ge \dots \ge \lambda_d > 0 $ the eigenvalues of $ \Sigma $
and $ \phi_{g, 1} \ge \dots \ge \phi_{g, d} > 0 $ the eigenvalues of $K_g$.

\subsection{Error bound for subspace recovery}
We investigate statistical efficiency of $g$-DPPCA in terms of the performance of subspace recovery under the elliptical model.
Let $ S = (X_1, \dots, X_n)  $ be a random sample from $ \mbox{EC}_d(\mu, \Sigma,\omega) $ where $\omega$ is a decreasing function.
Let $m \ll d$ be the number of leading eigenvectors we consider.
We denote $ \widetilde V_{g, m} \in \Oc(d, m) $ for the first $ m $ leading eigenvectors of $ \widetilde V_{g}(S) $ defined in Section \ref{sec:private-pca}.
Let $ U_m \in \Oc(d, m) $ be the first $ m $ eigenvectors of $ \Sigma $.

To derive how $ \col(\widetilde V_{g, m}) $ deviates from $ \col(U_m) $, we need a distance measure between two subspaces of $ \Rb^d $.
For this, we use the $\sin \Theta$ distance 
\citep{cai2018rateopt}. 
For $ V_1, V_2 \in \Oc(d, m) $, denote the smallest singular values of $ V_1^\top V_2 $ by $\varrho_m$. 
The largest principal angle between $ \mbox{col}(V_1) $ and $ \mbox{col}(V_2) $ is defined as  $ \Theta(\mbox{col}(V_1), \mbox{col}(V_2)) = \cos^{-1}(\varrho_m) $.
Then, the $\sin \Theta$ distance between two subspaces  $\mbox{col}(V_1)$ and $\mbox{col}(V_2)$ is defined as $\sin \Theta(\mbox{col}(V_1), \mbox{col}(V_2))$.
For example, if $ m=1 $, the $\sin \Theta$ distance is the sine of the angle between two vectors.

The following theorem describes how the privately estimated eigensubspace well approximates population subspace.

\begin{thm}\label{thm:utility}
    Let $ S = (X_1, \dots, X_n) $ be a random sample from $\mbox{EC}_d(\mu, \Sigma, \omega)$ with decreasing $\omega$.
    Let $ \beta \in (0, 1) $.
    If $ n \ge \frac{16}{3}(\|g\|_\infty^2/\phi_{g, 1} + 1) \log(2d/\beta) $, then  
\[
\Pb\left(
\sin \Theta \left(\col(\widetilde V_{g, m}(S)), \col(U_m)\right)
\lesssim  u_{g, m}
\right)
\ge 1-\beta,
\]
where 
\begin{align*}
    u_{g, m} = \frac{\phi_{g, 1}}{\phi_{g, m}- \phi_{g, m+1}}
    \left(\frac{\|g\|_{\infty}^2}{\phi_{g, 1}} \cdot \frac{\left(\sqrt{d} + \sqrt{\log(1/\beta)}\right)\sqrt{\log(1/\delta)}}{ n  \varepsilon} 
    + \sqrt{\frac{\|g\|_{\infty}^2}{\phi_{g, 1}} \cdot\frac{\log(d/\beta)}{n }}\right).
\end{align*}
Here, the probability is taken on both randomness of the additive mechanism itself and the data.
\end{thm}

To the best of our knowledge, Theorem \ref{thm:utility} is the first result that provides a subspace recovery error bound of the private PCA under the elliptical model.
As a special case, consider the non-private setting of $\varepsilon = \infty$, and $g = g_{sph}$ of \eqref{eq-gsph}, then Theorem \ref{thm:utility} recovers the upper bound appeared in \cite{han2018eca}; see Theorem 3.1 and Corollary 3.1 therein.
Thus, Theorem \ref{thm:utility} can be viewed as a generalization of the result of \cite{han2018eca}, which was originally stated under the non-private setting and the specific choice of $g = g_{sph}$, to a more general class of bounded $g$ and to the private setting.

The eigen-gap ratio $\phi_{g, 1} / (\phi_{g, m} - \phi_{g, m+1})$ in the upper bound $u_{g, m}$ arises due to Davis-Kahan type bound \citep{yu2015useful, Han2025winsor}.
The first term of $u_{g, m}$ is owing to the privacy constraints, 
and the second term is resulting from the sample estimation efficiency of $\widehat K_g$.
We note that $u_{g, m}$ depends on the specific choice of $g$ through $\phi_{g, i}$ and $\|g\|_\infty$.
Since what describes the underlying model is $\Sigma$ (not  $K_g$), it is desirable to express the upper bound using the model parameter $\Sigma$. 
Such expression should be obtained for each particular choice of $g$,
and we study the two examples of spherical transformation and winsorization in Section \ref{subsec:two-g}.

\subsection{Robustness against data corruption} \label{sec-theory-robust}

The results on the error bound of $ g $-DPPCA presented in the previous subsection are valid for heavy-tailed elliptical distributions even without finite moments.
This implies robustness, as the performance of $g$-DPPCA is guaranteed under heavy-tailed conditions. In this section, we further demonstrate robustness of the proposed method under arbitrary data corruption by conducting the classical breakdown point analysis and subspace recovery analysis.


We begin with a breakdown point analysis \citep{hampel1986robust, huber2011robust} of $g$-DPPCA.
The breakdown point of a statistic is the smallest fraction of corrupted samples in $ n $ observations that can cause the statistic to yield arbitrarily large errors.
For example, the sample mean on $\mathbb{R}$ has a breakdown point of $1/n$, while the sample median achieves $ \lfloor (n+1)/(2n) \rfloor $. Hence, the median with larger breakdown point is more robust than the mean. 

Since we focus on the problem of estimating the subspace spanned by the first few eigenvectors of $ \Sigma $, we adopt the notion of breakdown point for subspace-valued statistics, introduced in \cite{Han2025winsor}.
Denote the Grassmannian manifold by $ \mbox{Gr}(d, m) $, the set of $ m $-dimensional linear subspaces of $ \mathbb{R}^d $.

\begin{defn}[Breakdown point of subspace-valued statistic] \label{def:bp}
    For $ \mathcal{V}: (\Rb^d)^n \to \mbox{Gr}(d, m)  $, the breakdown point of $ \mathcal{V} $ at 
    $ S = (x_1, \dots, x_n) \in  (\Rb^d)^n $ is defined as 
    \begin{equation*}
        \mbox{bp}(\mathcal{V}; S) = \min_{1 \le r \le n} \left\{
            \frac{r}{n} : \sup_{S_{r/n}} \Theta(\mathcal{V}(S_{r/n}), \mathcal{V}(S)) = \frac{\pi}{2}
        \right\}.
    \end{equation*}
    Here, the supremum takes all possible $ r/n $ corrupted dataset $ S_{r/n} $ which has the form of $(x_1, \dots, x_{n-r}, y_1, \dots, y_r)$ for some $y_1, \dots, y_r \in \Rb^d.$
\end{defn}

For a dataset $ S = (x_1, \dots, x_n) \in (\mathbb{R}^d)^n $, 
let $ \widehat V_m (S) \in \Oc(d, m)$ be the matrix of first $m$ eigenvectors of 
$ \widehat K_g(S) $.
Define the subspace-valued statistic $ \Vc_m^g $ by $ \Vc_m^g(S) = \col(\widehat V_m(S)) $.
Then the following theorem shows that the breakdown point of $ \Vc_m^g $ is bounded below.

\begin{thm} \label{thm:bp-Kendalltau}
    Fix a dataset $ S = (x_1, \dots, x_n) \in (\mathbb{R}^d)^n $.
    Then, the breakdown point of $ \Vc_m^g $ at $ S $ is bounded below as
    \begin{equation*}
        \textnormal{\mbox{bp}}(\mathcal{V}_{m}^g; S) 
        \ge \frac{\hat \phi_{g, m} - \hat \phi_{g, m+1}}{8 \|g\|_{\infty}^2},
    \end{equation*}
    where $ \hat \phi_{g, i} $ is the $ i $th eigenvalue of $ \widehat K_g(S) $.
\end{thm}

Consider conducting PCA via sample covariance matrix $ \widehat \Sigma(S) = n^{-1} \sum_{i=1}^n x_i x_i^\top $.
Let $ \Vc_m $ be the corresponding PC subspace-valued statistic.
Then, it holds that $ \mbox{bp}(\mathcal{V}_m; S) = 1 / n $ \citep{Han2025winsor} and vanishes to $ 0 $ as $ n \to \infty $.
In contrast, the lower bound in Theorem \ref{thm:bp-Kendalltau} converges to 
$ (\phi_{g, m} - \phi_{g, m+1}) / (8\|g\|_{\infty}^2) > 0$ as $ n \to \infty $.
This shows that $ g $-DPPCA in the non-private setting ($ \varepsilon = \infty $) is more robust than the standard sample covariance-based estimator in terms of breakdown point.

The above argument on the non-private $\widehat K_g$ sheds light on the robustness of $g$-DPPCA. To see this, recall that $ \widetilde K_g(S) = \widehat K_g(S) + \xi $, where $\xi$ denotes an $ (\varepsilon, \delta) $-DP calibrated Gaussian noise, and that $g$-DPPCA, $ \widetilde \Vc_m^g(S)$, returns the dimension-$m$ eigenspace of $\widetilde K_g(S)$. Since $\xi$ is independent of $\widehat K_g$, if the non-private $\Vc_m^g$ does not break down, then the privatized $\Vc_m^g$ also avoids breakdown. This implies that $ \widetilde \Vc_m^g(S)$ is also robust to arbitrary data contamination. We note, however, that the notion of breakdown in Definition \ref{def:bp} is, not well-defined for randomized mechanisms; a formal extension of the breakdown point to randomized mechanisms is left for future work.


%
%

Next, we investigate the error bound on subspace recovery under data contamination to measure the robustness of $g$-DPPCA in terms of utility.
This is a common approach for analyzing the robustness of differentially private methods (see, e.g.,  \cite{liu2022robustdp}).
The following theorem provides such a bound for elliptically distributed data where a fraction of the samples is arbitrarily corrupted.
We note that the theorem is valid for both  private and non-private settings.

\begin{thm} \label{thm:corrupted}
    Let $ S = (X_1, \dots, X_n) $ be a random sample from $\mbox{EC}_d(\mu, \Sigma, \omega)$ with decreasing $\omega$.
    Denote an $ \alpha $-fraction corrupted dataset as 
    $S_\alpha = (X_1, \dots, X_{n(1-\alpha)}, y_1, \dots, y_{n\alpha})$ 
    where $ (y_1, \dots, y_{n\alpha}) \in (\Rb^d)^{n \alpha} $ is an arbitrarily corrupted part of $ S $.
    Let $ \beta \in (0, 1) $.
    If $ n \ge \frac{16}{3(1-\alpha)}(\|g\|_{\infty}^2/\phi_{g, 1} + 1) \log(2d/\beta) $, then with 
    probability greater than $ 1- \beta $, 
    \begin{equation*}
       \sup_{y_1, \dots, y_{n\alpha}} \sin \Theta \left(\col(\widetilde V_{g, m}(S_{\alpha})), \col(U_m)\right)   
    \lesssim \frac{\phi_{g, 1}}{\phi_{g, m} - \phi_{g, m+1}}\left(I_1 + I_2 + I_3\right),
    \end{equation*}
    for some absolute constant $ C > 0 $, where
    \begin{align*}
        I_1 &= \frac{\|g\|_{\infty}^2}{\phi_{g, 1}} \frac{\left(\sqrt{d} + \sqrt{\log(1/\beta)}\right)\sqrt{\log(1/\delta)}}{n \varepsilon}, \\
        I_2 &= 
        \sqrt{\frac{(\|g\|_{\infty}^2/\phi_{g, 1} + 1) \log(d/\beta)}{(1-\alpha)n}}, \\
        I_3 &=
        \alpha(2-\alpha) \left(1 + \frac{\|g\|_{\infty}^2}{\phi_{g, 1}} \right).
    \end{align*}
\end{thm}

The first term $ I_1 $ is due to the privacy constraint and is independent of data contamination. 
This term vanishes in the non-private setting, i.e., as $\varepsilon \to \infty$, and is in fact the same as the first term of $u_{g, m}$ in Theorem \ref{thm:utility}.
The terms $ I_2 $ and $ I_3 $ originate from data corruption and are unrelated to privacy constraints.
When $ \alpha \to 0 $, $ I_2 $ converges to the second term of $u_{g, m}$ while $ I_3$ vanishes, recovering 
the bound of Theorem \ref{thm:utility}.
In contrast, if $ \alpha \in (0, 1) $  is fixed, then $ I_3 $ does not vanish even in the large-sample asymptoticss, i.e., $ n \to \infty $. 
Nevertheless, whenever $\alpha \to 0$ as $n \to \infty$, all three terms of the error bound becomes zero. For instance, when $\alpha = 1/\sqrt{n}$, the error bound becomes zero  as $n$ increases, while the number of corrupted samples $n\alpha = \sqrt{n}$ still diverges. Therefore, with a modest level of data corruption, $g$-DPPCA remains robust to contamination while satisfying differential privacy.



\subsection{Two examples of \textit{g}-DPPCA}\label{subsec:two-g}

In this subsection, we express the upper bound $u_{g, m}$ in Theorem \ref{thm:utility} in terms of the eigenvalues $\{\lambda_j\}_{j=1}^d $ of $\Sigma$, for two specific choices of $g$:
the spherical transformation $g_{sph}$, defined in \eqref{eq-gsph}, and the winsorization $ g_{wins}^{(r)} $, defined in \eqref{eq-gwins}.
We use the notation $K_{sph}$ for $K_g$ when $ g = g_{sph} $, and $K_{wins}^{(r)}$ when $g =g_{wins}^{(r)}$.
Write $\phi_{sph, \ell}$ for the $\ell$th largest eigenvalue of $ K_{sph}$
and $\phi_{w, \ell}^{(r)}$ for the $\ell$th largest eigenvalue of $K_{wins}^{(r)} $.

First, consider the spherical transformation. 
Under the elliptical distributional assumption, the eigenvalues of $ K_{sph} $ can be expressed 
as the expectation of a self-normalized weighted quadratic form of a Gaussian random vector: For any $ 1 \le \ell \le d $, 
\begin{equation*} 
    \phi_{sph, \ell}
    = \Eb \left[\frac{\lambda_\ell Y_\ell^2}{\sum_{j=1}^d \lambda_j Y_j^2}\right],
\end{equation*}
where $(Y_1, \dots, Y_d)^\top \sim N_d(0, I_d).$
This result is well known in the context of spherical PCA; for example, see \cite{croux2002sign} and \cite{han2018eca}. 

The following proposition describes how the error bound $u_{sph, m}$ in Theorem \ref{thm:utility}
can be written in terms of $\phi_{sph, \ell}$, $n, d$ and $\varepsilon$.
Furthermore, based on the recent result on the bound of $ \phi_{sph, \ell} $ \citep{han2018eca},  an upper bound of $u_{sph, m}$ can be written in terms of eigenvalues of $\Sigma$.
Let $ \er(\Sigma) =  \tr(\Sigma)/ \lambda_1$ be the effective rank of $ \Sigma $.

\begin{prop}\label{prop:errbound-sph}
    The upper bound in Theorem \ref{thm:utility} for spherical transformation becomes 
    \[
        u_{sph, m} \asymp 
        \frac{1}{\phi_{sph, m} - \phi_{sph, m+1}}
        \left(
            \frac{\sqrt{d}}{n \varepsilon} +
            \sqrt{\frac{\phi_{sph, 1} \log(d)}{n}}
        \right).
    \]
    Here, the logarithmic terms of $\beta$ and $\delta$ are ignored.
    In addition, suppose that 
    $\|\Sigma\|_F \log d = \tr (\Sigma) \cdot o(1)$ as $d$ increases.
    Then, we have
    \begin{equation}\label{eq:sph-upperbound}
        u_{sph, m}
        \lesssim 
            \frac{\lambda_1}{\lambda_m - \lambda_{m+1}} 
            \left(\frac{\er(\Sigma)\sqrt{d}}{n \varepsilon} + 
            \sqrt{\frac{\er(\Sigma) \log(d)}{n}}\right).
    \end{equation}
\end{prop}

The condition $\|\Sigma\|_F \log d / \tr(\Sigma) = o(1)$ is a mild regularity condition, permitting the first few leading eigenvalues to increase with the dimension $d$.
For example, suppose $\Sigma$ has a $m$-spiked covariance structure of $\Sigma = \lambda U_m U_m^\top + \sigma^2 I_d$ with signal to noise ratio as $\lambda / \sigma^2 \asymp d^{a}$ for some $a \in [0, 1),$ then this condition is satisfied.
(For the proof, see Appendix \ref{subsec-app-c1}.)

Next, consider the winsorization transformation with radius $r > 0$. 
For any $ 1 \le  \ell \le d  $, we can show that
\begin{equation} \label{eq:eigval-wins}
\phi_{w, \ell}^{(r)} 
= \Eb\left[\min\left(R^2,\: \frac{r^2}{\sum_{j=1}^d \lambda_j S_j^2}\right) \lambda_{\ell}  S_{\ell}^2 \right],
\end{equation}
where $R^2 \overset{d}{=} \frac{1}{2} (X- \widetilde X)^\top \Sigma^{-1} (X- \widetilde X)$ and $(S_1, \dots, S_d)  \sim \mbox{Unif}(\Sb^{d-1})$ are independent. 
Here,  $\widetilde X$ is an independent copy of $X$.
We defer the proof of \eqref{eq:eigval-wins} to Appendix \ref{app-proof}.

As in the case of spherical transformation, an upper bound $u_{wins, m}^{(r)} $ can be written in terms of
$\phi_{w, \ell}^{(r)} $ and the radius $r$.


\begin{prop} \label{prop:errbound-wins}
    For any $ r > 0 $, the upper bound in Theorem \ref{thm:utility} for
    winsorization becomes
    \begin{equation} \label{eq:wins_asymp}
        u_{wins, m}^{(r)}
        \asymp \frac{r^2}{\phi_{w, m}^{(r)} - \phi_{w, m+1}^{(r)}}
        \left(
            \frac{\sqrt{d}}{n \varepsilon} 
            + \sqrt{\frac{\phi_{w, 1}^{(r)}}{r^2} \cdot \frac{\log (d)}{n}}
        \right).
    \end{equation}
    Furthermore, if the radius $r$ is chosen to satisfy 
    \begin{equation} \label{eq-winscond}
        \Pb(R^2 \ge r^2/\lambda_d) \asymp 1,
    \end{equation}
    it holds that $u_{wins, m}^{(r)} \asymp u_{sph, m}.$
\end{prop}

The choice of $r$ affects the performance of $g_{wins}^{(r)}$-DPPCA.
If we take $r \to \infty$, then $u_{wins, m}^{(r)} \to \infty$ holds, which can be deduced from \eqref{eq:wins_asymp}.
On the other hand, with suitably chosen radius that satisfies \eqref{eq-winscond}, the effect of the winsorization becomes similar to the spherical transformation.
For both Gaussian and multivariate $t$-distributions, which are typical examples of light-tailed and heavy-tailed elliptical distribution, respectively, 
choosing $ r \lesssim \sqrt{d \lambda_d} $ is sufficient for condition \eqref{eq-winscond}; see Appendix \ref{subsec-app-c2} for details. 

In practice, one can simply set  $r = c\sqrt{d}$, where $c \approx \lambda_d$ is determined based on prior or publicly available information, or alternatively choose $r$ as the privatized first quartile of the paired differences $\|X_i - X_j\|_2/\sqrt{2}$ using a small privacy budget; see Appendix~\ref{app-sim-radius} for a more detailed discussion on the practical choice of $r$.


\begin{remark}\label{remark_3-1} 
We emphasize that our error bound \eqref{eq:sph-upperbound} is derived under a broad class of elliptical distributions. To put the result into perspective, we first compare it with the error bound of \textit{Analyze Gauss} \citep[see Section \ref{sec:competingmethods},][]{dwork2014analyze}.
Since $ \er(\Sigma) \le d $, our bound \eqref{eq:sph-upperbound} is lower than the  bound of Analyze Gauss
$O\left(\frac{\lambda_1}{\lambda_m - \lambda_{m+1}} \left(
                \frac{d^{3/2}}{\varepsilon n} + \sqrt{\frac{d}{n}} 
        \right)
    \right)$ \citep{liu2022dp}. 
We next compare the bound with that of a recently proposed \textit{SGPCA}  \citep{cai2024optimal}, which is obtained under strong moment and structural assumptions, including sub-Gaussianity and spiked covariance models. Assuming an $m$-spike model $\Sigma = \lambda U_mU_m^\top + \sigma^2 I_d$ with signal-to-noise ratio $\lambda / \sigma^2 \asymp d^{a}$, where $a \in [0, 1)$, our error bound simplifies to 
    \begin{equation} \label{rmk3.1-sph}
    O\left(
    \left(1 + \frac{\sigma^2}{\lambda} \right)
        \left(\frac{\mbox{er}(\Sigma)\sqrt{d}}{\varepsilon n} + \sqrt{\frac{\mbox{er}(\Sigma)}{n}}\right) 
    \right) =  O\left( \frac{d^{\frac{3}{2}-a}}{n\varepsilon} + \sqrt{\frac{d^{1-a}}{n}}  \right),
    \end{equation}
while the bound of SGPCA becomes
    \begin{equation} \label{rmk3.1-cai}
        O\left(
    \left(\frac{\sigma^2}{\lambda} + \sqrt{\frac{\sigma^2}{\lambda}} \right)
        \left(\frac{d}{\varepsilon n} + \sqrt{\frac{d}{n}}\right) 
    \right) = O\left(\frac{d^{1-\frac{a}{2}}}{n\varepsilon} + \sqrt{\frac{d^{1-a}}{n}} \right).
    \end{equation}
Comparing the privacy-related first terms of (\ref{rmk3.1-sph}) and (\ref{rmk3.1-cai}), we observe that the error term of SGPCA is smaller by a factor of $d^{\frac{1-a}{2}}$.
While this gap narrows as the magnitude of spikes increases ($a \to 1$), it should be noted that the error bound of \cite{cai2024optimal} is derived under a potentially restrictive sub-Gaussian assumption.
 \end{remark}

\section{Numerical studies} \label{sec:numerical-study}

In this section,  we evaluate the empirical performances of our proposed methods, 
specifically for $g_{sph}$-DPPCA and $g_{wins}^{(r)}$-DPPCA, using both simulated and real datasets. 
For winsorization, we set $ r = \sqrt{d} $ in the whole numerical study.
The utility is measured by how well a given methods recovers the underlying population eigensubspace.

\subsection{Competing methods}\label{sec:competingmethods}
We consider three competing methods,  proposed by \cite{dwork2014analyze},  \cite{maunu2022stochastic}, and \cite{cai2024optimal}.
Many other works on private PCA exist, but most of them do not fit our settings or are challenging to implement because their algorithms are not polynomial-time executable, or highly dependent on the tuning procedure and population parameters.
The above three methods are implementable, and the detailed algorithms used in our numerical studies are provided in Appendix \ref{app-C}.

The analyze Gauss mechanism (AG), proposed by \cite{dwork2014analyze}, adds Gaussian noise to the sample covariance matrix, and then extracts eigenvectors from the noisy sample covariance matrix.
\cite{cai2024optimal}'s proposal is called the spiked covariance Gaussian PCA (SGPCA) which assumes a spiked covariance model and sub-Gaussianity of data distribution. 
SGPCA applies Gaussian mechanism to $ \widehat V_m \widehat V_m^\top $ and then takes eigenvectors from the noisy projection matrix, where $\widehat V_m$ is the matrix of the first $m$ eigenvectors of the sample covariance matrix.
Finally, the private PCA method of \cite{maunu2022stochastic}, called noisy stochastic geodesic gradient descent (NSGGD) solves an $L_1$-PCA optimization problem by iterating over the (approximate) geodesic gradient in the Stiefel manifold, which consists of orthogonal matrices in $ \Oc(d, m) $.
In contrast to AG and SGPCA,  NSGGD is designed to be robust to outliers.

\subsection{Simulation study}  \label{sec:sim-study}

To evaluate the empirical performance of the proposed methods, we generate a simulated dataset from elliptical distributions and fit the private estimators for PC directions.
We consider two types of elliptical distributions with a dispersion matrix $\Sigma$ possessing two-spiked structure. 
For a given data dimension $d$, we set 
$\Sigma = (\lambda_1 - \lambda_d)v_1v_1^\top + (\lambda_2 - \lambda_d)v_2 v_2^\top + \lambda_d I_d,$
where $v_1$ and $ v_2$ are the first two eigenvectors and $\lambda_1 > \lambda_2 > \lambda_d = \dots = \lambda_d$ are the eigenvalues of $\Sigma$.
We generate a dataset of size $N$ from  
either the Gaussian distribution $N(0, \Sigma)$ or the centered multivariate $t$-distribution with degree of freedom 1, $ t_1(\Sigma) $, i.e., $t_1(\Sigma) \overset{d}{=} \Sigma^{1/2}Z/\sqrt{\chi^2_1}$ where $Z \sim N_d(0, I_d)$, and $\chi^2_1$ is independent with $Z$.
We note that $t_1(\Sigma)$, also known as the multivariate Cauchy, does not have a finite mean.
Also, to investigate the robustness of comparing methods, we sample a dataset from a 5\% contaminated Gaussian distribution as follows.
First, sample dataset from $N(0, \Sigma)$. 
Then, replace randomly selected 5\% of the data by samples from $N(c_{\perp} v_{\perp}, 0.05^2 I_d)$, where $v_{\perp} \in \Rb^d$ is an unit vector orthogonal to $v_1$ and $v_2$ and $c_{\perp} > 0$ is a constant.
In the simulation, we set the eigenvalues as $ (\lambda_1, \lambda_2, \lambda_d) = (10, 5, 1) $, and the eigenvectors as $v_1 = (1, 1, 1, 1, 0_{d-4})^\top/2 $ and $v_2 = (1, -1, 1, -1, 0_{d-4})^\top/2 $.
For the contaminated Gaussian, we use $v_{\perp} = (0, 1, 0, -1, 0_{d-4})^\top/\sqrt{2}$ and $c_{\perp} = 2.5\lambda_1$.

\begin{figure}[t]
\centering
\includegraphics[width=.9\textwidth]{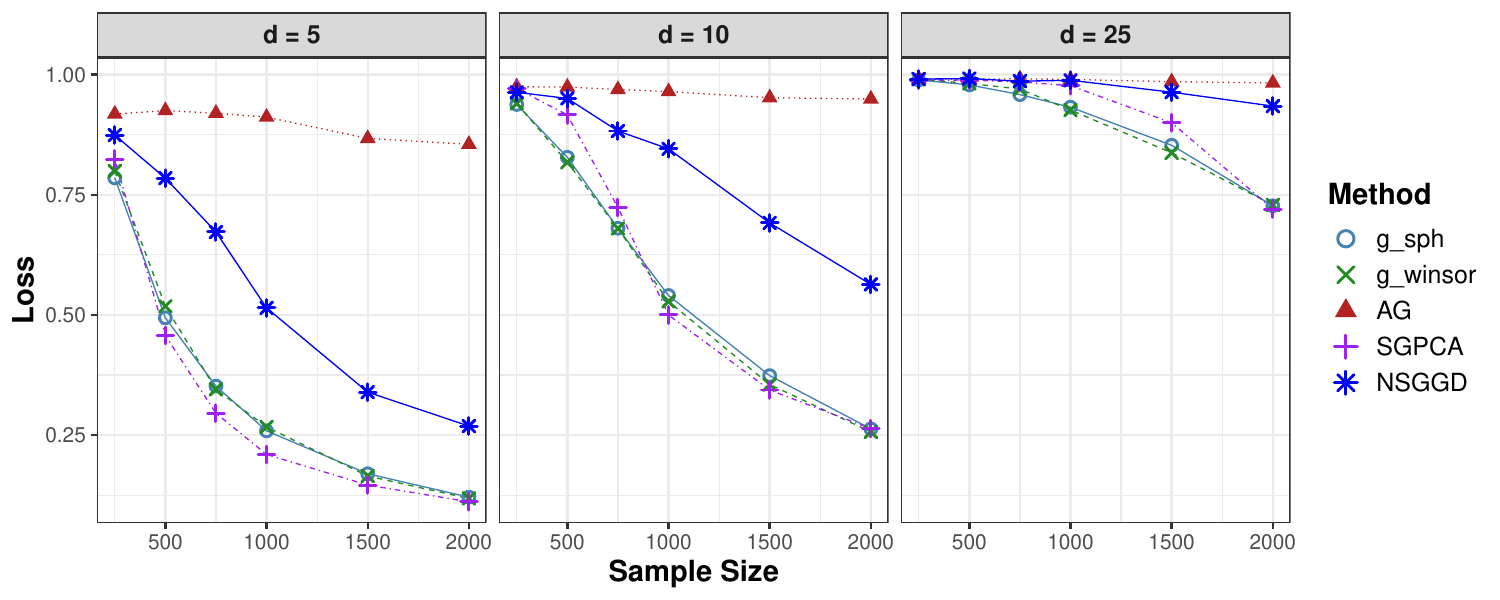}
\caption{Average $\sin \Theta$ losses over increasing sample size and dimension under the Gaussian distribution $N(0, \Sigma)$ with fixed privacy parameters $ (\varepsilon, \delta)= (0.5, 10^{-5}) $.}
\label{fig:sim_Gaussian}
\end{figure}

\begin{figure}
\centering
\includegraphics[width=.9\textwidth]{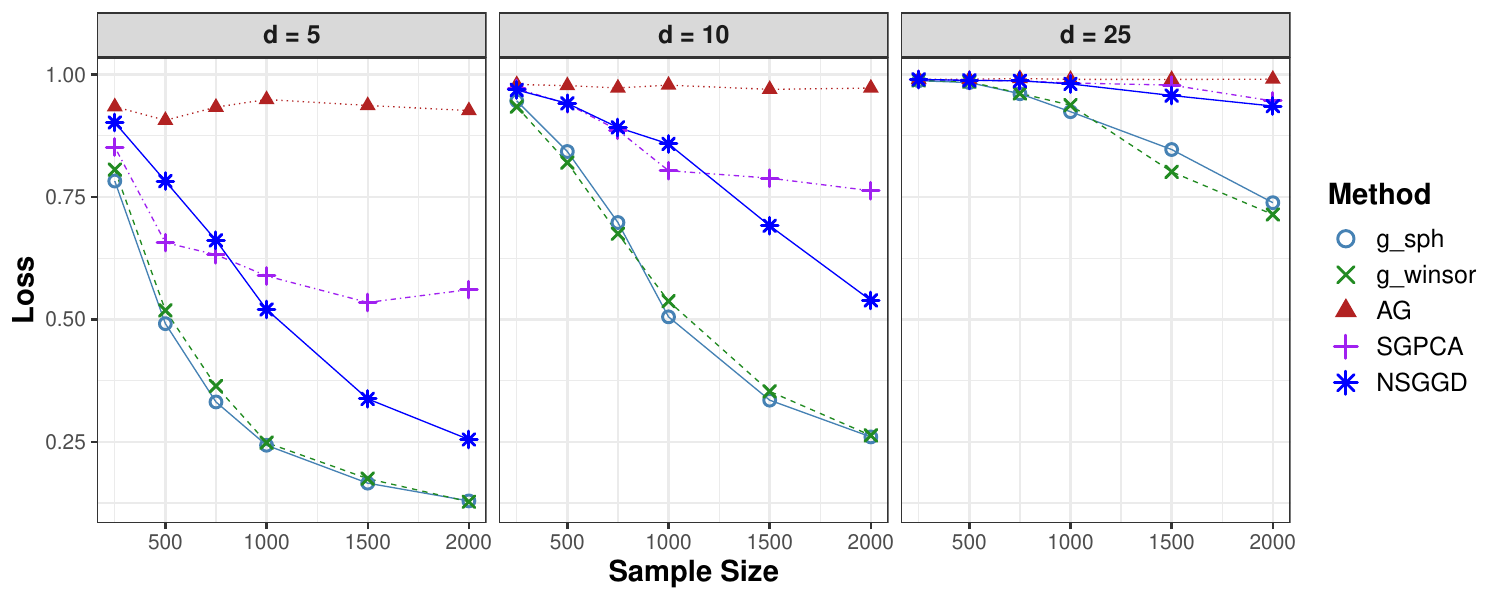}
\caption{Average $\sin \Theta$ losses over increasing sample size and dimension under the multivariate $t$-distribution $t_1(\Sigma)$ with fixed $ (\varepsilon, \delta)= (0.5, 10^{-5}) $.}
\label{fig:sim_t1}
\end{figure}

\begin{figure}
\centering
\includegraphics[width=.9\textwidth]{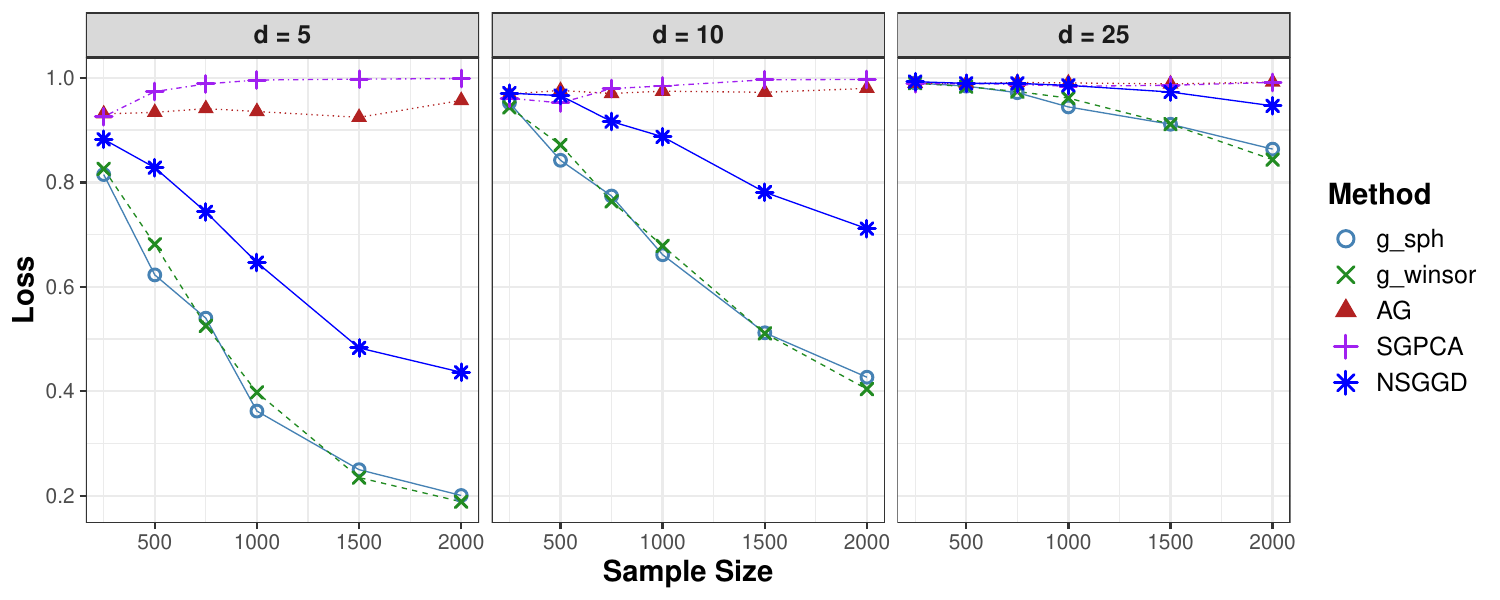}
\caption{Average $\sin \Theta$ losses over increasing sample size and dimension under 5\% contaminated Gaussian distribution with fixed $ (\varepsilon, \delta)= (0.5, 10^{-5}) $.}
\label{fig:sim_cg05}
\end{figure}

First, we compare the performance of the methods under a fixed privacy budget.
For each setting, we obtain the first two privately estimated PCs $\widetilde V_2 = [\tilde v_1, \tilde v_2]$ and measure the performance by  $\sin \Theta(\col(\widetilde V_2), \col(V_2))$, where $ V_2 = [v_1, v_2] $.
The sample size ranges from $ n \in \{250, 500, 750, 1000, 1500, 2000\} $, and the dimension $ d $ ranges from $ d \in \{5, 10, 25\} $.
The privacy parameters are fixed as $ (\varepsilon, \delta) = (0.5, 10^{-5}) $. 
We repeat this procedure 100 times for each $ (n, d) $ pair, and the averaged losses are reported in Figures \ref{fig:sim_Gaussian}--\ref{fig:sim_cg05} according to the distributional settings.

The simulation results for the Gaussian distribution are presented in Figure \ref{fig:sim_Gaussian}.
The proposed $g_{sph}$- and $g_{wins}$-DPPCA mechanisms perform comparably to each other and outperform AG and NSGGD.
However, SGPCA shows better performance compared to our methods when $ d = 5 $.
This result is expected, as SGPCA is theoretically (near) optimal private PCA method under the Gaussian assumption with spiked covariance matrix structure. 
Furthermore, we assume that the eigenvalues of $ \Sigma $ are known a priori when implementing SGPCA, which is highly favorable to SGPCA. See Algorithm \ref{alg-sgpca} in Appendix \ref{app-C} for details.
However, the performance gap between SGPCA and our methods diminishes as the sample size increases.

Figure \ref{fig:sim_t1} presents the simulation results for $t_1(\Sigma)$, which has a heavier tail than the Gaussian distribution. 
Notably, the proposed methods
outperform all other methods.
In contrast to the Gaussian case, NSGGD outperforms SGPCA as the sample size increases.
This is due to the robustness of NSGGD, while SGPCA does not guarantee any robustness beyond Gaussian data.
Nevertheless, our methods outperform NSGGD in all cases, indicating that $g$-DPPCA is more efficient and robust than NSGGD under heavy-tailed distribution.
It remains unclear which of the two, $ {g}_{sph} $ and ${g}_{wins} $, performs better.

For the third case of data contamination, Figure \ref{fig:sim_cg05} shows similar patterns to those in Figure \ref{fig:sim_t1}.
Our proposed methods are robust to data contamination since the losses decrease as the sample size increases.
The losses of AG and SGPCA, however, do not decrease as the sample size grows.
This is because both methods rely on the sample covariance estimator, which is highly sensitive to data contamination.
Since NSGGD also shows robustness, its loss decreases consistently across all cases.
Nevertheless, our methods outperform NSGGD by a considerable margin.

\begin{figure}[t]
\centering
\includegraphics[width=.9\textwidth]{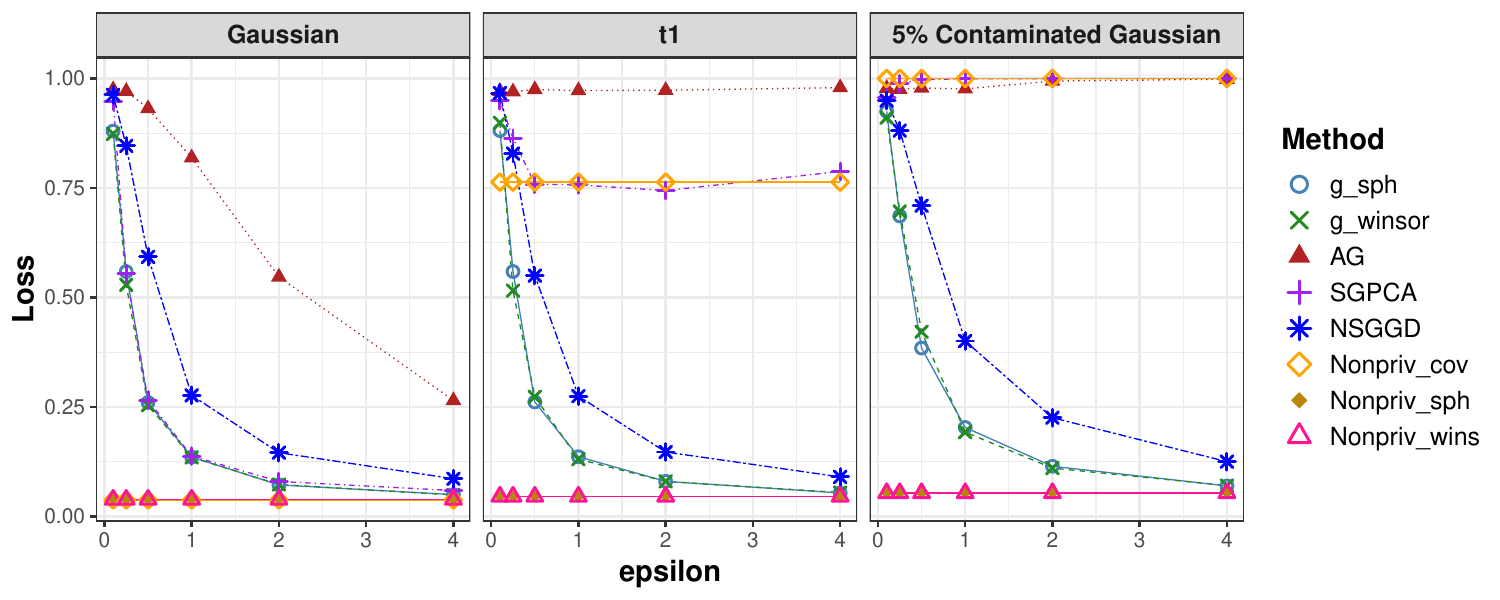}
\caption{Comparison of $\sin \Theta$ losses as the privacy parameter $\varepsilon$  increases under three types of distributions with fixed $ n=2000, d=10$ and $\delta=10^{-5}$.}
\label{fig:sim_eps}
\end{figure}

Next, we investigate the performance of all methods with varying privacy parameter $\varepsilon$ while the sample size and dimension are fixed as $ (n, d) = (2000, 10) $.
We set $\delta = 10^{-5}$ and consider $ \varepsilon \in [0.1, 4] $.
The simulation results are summarized in Figure \ref{fig:sim_eps}, which also compares with the results from non-private mechanisms.
Once again, similar patterns emerge: $ g_{sph} $ and $ {g}_{wins} $-DPPCA provide the
best performance compared to the other methods across all cases of distributions except for the Gaussian case. 
In the cases of $ t_1 $ and contaminated Gaussian, SGPCA performs worse than $ g $-DPPCA methods and NSGGD.
The simulated loss of NSGGD decreases as $ \varepsilon $ increases; however, the performance gap between our methods and NSGGD does not narrow down to zero.

Additional numerical studies on robustness are provided in the appendix.
In Appendix \ref{app-sim-contam}, we investigate contaminated Gaussian data by varying $c_{\perp}$ and $\rho$, while Appendix \ref{app-sim-nonelliptic} presents simulation results for non-elliptical data.
Together, these results demonstrate that our proposed $g$-DPPCA methods are more robust to contamination and distributional deviations than competing methods.

\subsection{Visualization: Europe map data}

\begin{figure*}[t]
    \centering
    \begin{subfigure}[b]{0.32\textwidth}
        \centering
        \includegraphics[width=\textwidth]{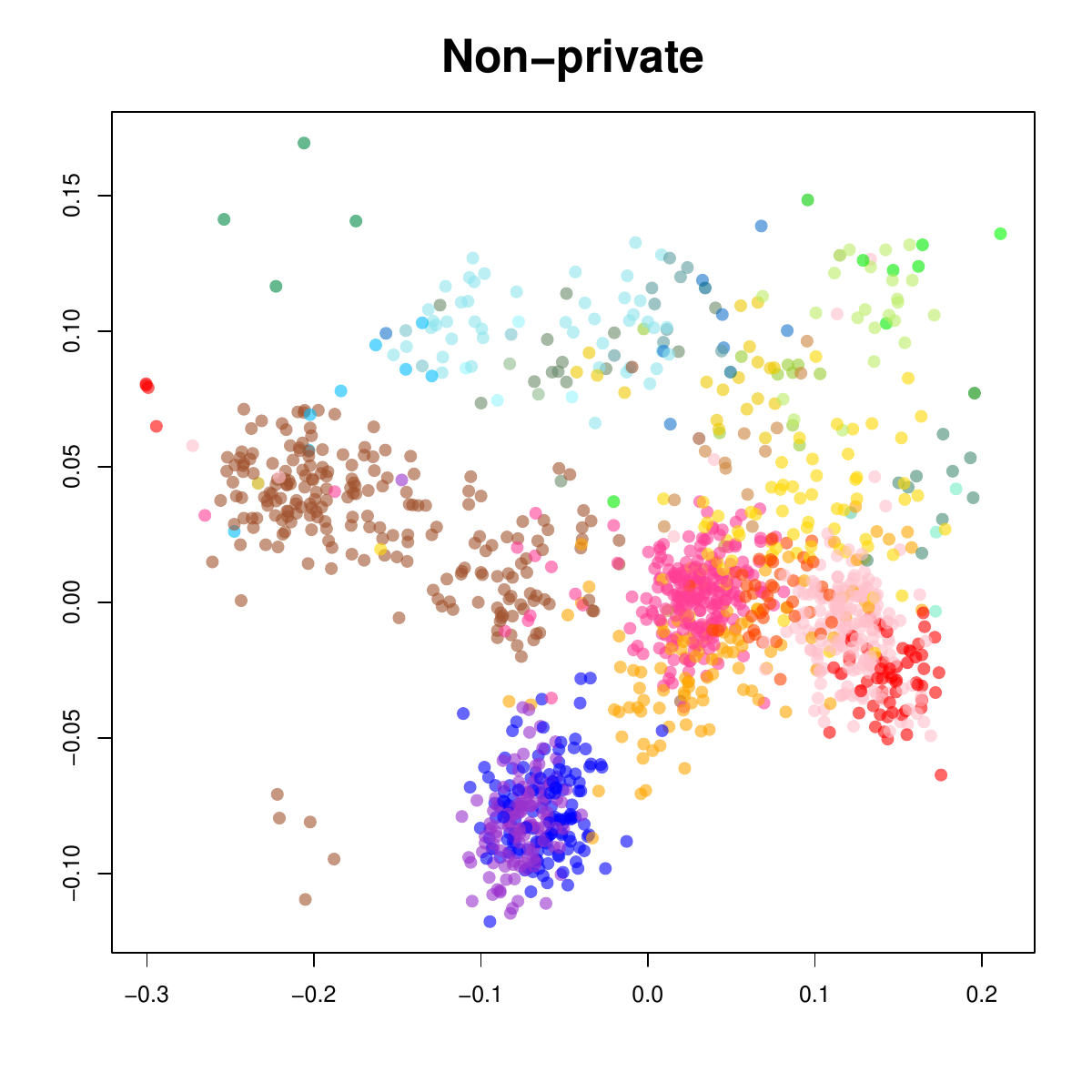}
    \end{subfigure}
    \begin{subfigure}[b]{0.32\textwidth}
        \centering
        \includegraphics[width=\textwidth]{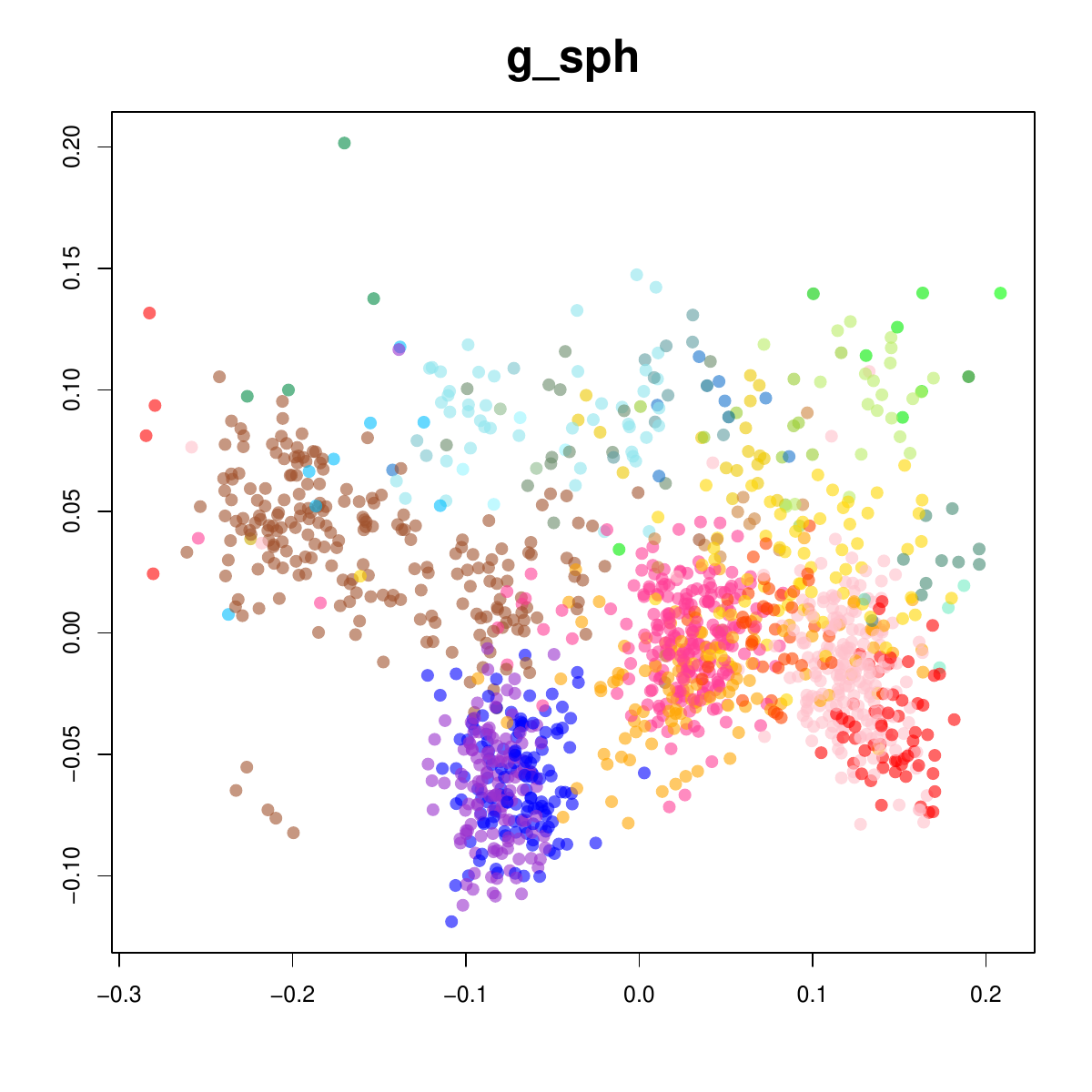}
    \end{subfigure}
    \begin{subfigure}[b]{0.32\textwidth}
        \centering
        \includegraphics[width=\textwidth]{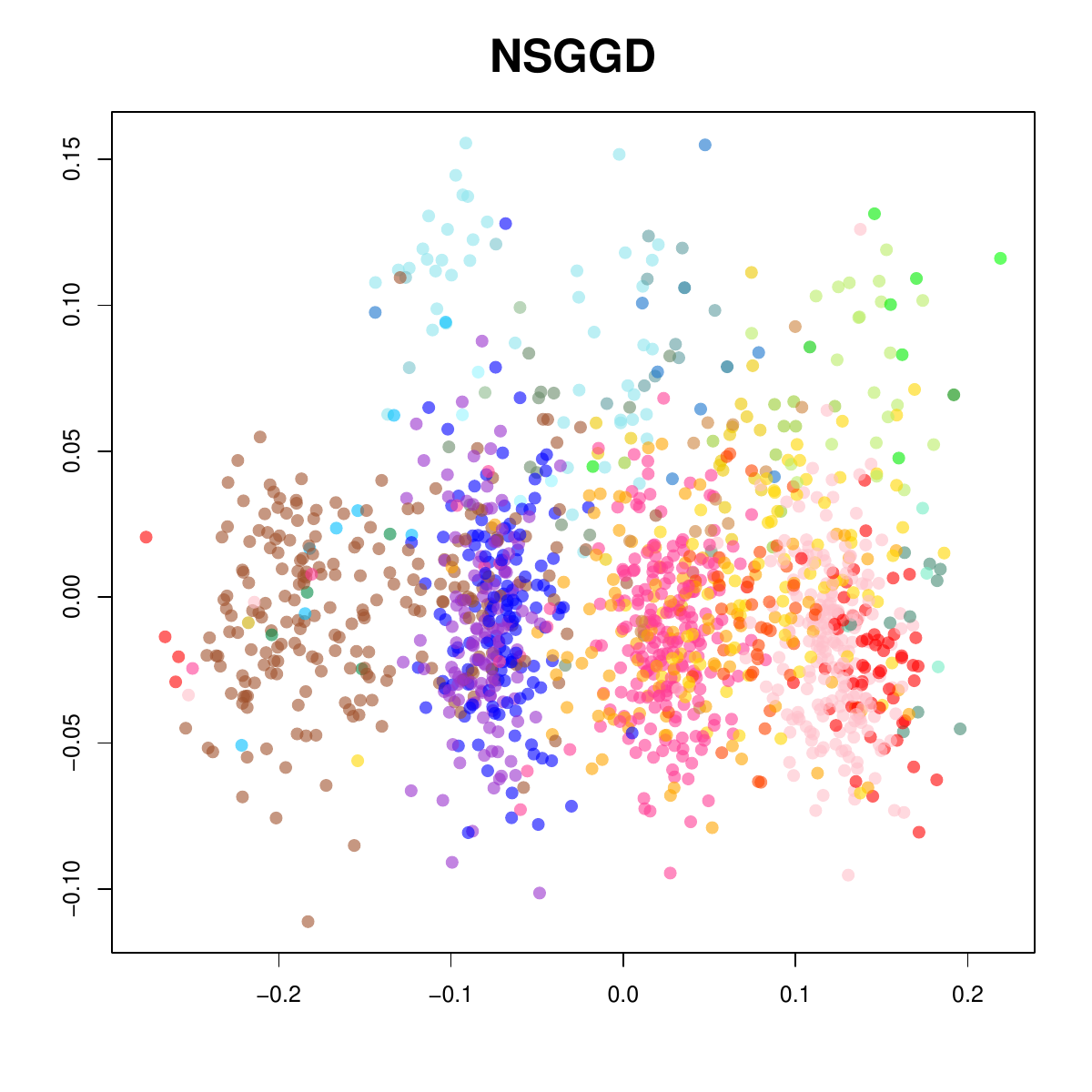}
    \end{subfigure}
    \caption{Projection maps of the genetic data of European individuals.}
    \label{fig:eur-maps}
\end{figure*}

The most common use of PCA is visualizing high-dimensional data by projecting it onto a low-dimensional subspace. 
To evaluate the practical utility of private PCA methods, we apply them to the genetic data of 1,387 European individuals originally studied by \cite{novembre2008genes}. 
In that study, a two-dimensional projection of the genetic data via traditional PCA closely resembled the geographic layout of European countries.

While the original high-dimensional genotype dataset is not publicly available due to privacy concerns, a dimension-reduced version is publicly accessible. 
We obtain a 20-dimensional version of the dataset by following the process outlined by  \cite{biswas2020coinpress}. 
We then compute first two leading principal components with $ (\varepsilon, \delta) = (2, 10^{-4}) $.
The dataset is subsequently projected onto the subspace spanned by these two private principal components, and each individual is colored according to their nationality. 

In Figure \ref{fig:eur-maps}, we provide projection plots of $ g_{sph} $-DPPCA and NSGGD for comparison.
We also include the projection plot obtained from non-private PCA utilizing the sample covariance estimator for a baseline comparison.
We do not present SGPCA since it either requires prior knowledge of the population eigenvalues or an additional privacy-preserving estimation step, which makes the problem more complex, and also omit $g_{wins}$-DPPCA since its performance is similar to that of $ g_{sph} $-DPPCA. 

We observe that ${g}_{sph}$-DPPCA produces a projection map that closely resembles the non-private PCA projection, effectively capturing the underlying geographic structure in the data.
In contrast, the projection map from NSGGD appears less informative. 
On the first PC direction ($x$-axis), NSGGD well reflects the variation with respect to countries, but its separability between nationalities decreases in the second PC direction. 

We note that the individual scores in the projection plots are not differentially private since the raw data were used without privatization. Nevertheless, these projection plots are valuable  because they illustrate how the proposed methods effectively capture the underlying geographic structure of genetic data.

\bibliographystyle{asa}
\bibliography{library}

\newpage
\appendix

\section{Omitted proofs} \label{app-proof}
In the proofs, a variant of Davis-Kahan sin $\Theta$ theorem is frequently used.
\begin{lem}\label{lem:kahan}[Lemma 10 of \cite{Han2025winsor}, originally from \cite{yu2015useful}]
    Let $\Sigma, \widehat{\Sigma} \in \Rb^{p\times p}$ be symmetric, with eigenvalues $\lambda_1 \geq \dots \geq \lambda_p$ and $\widehat{\lambda}_1 \geq \dots \geq \widehat{\lambda}_p$, respectively. Assume that $\lambda_d > \lambda_{d+1}$. Let $\Vv = (\vv_1,\dots,\vv_d) \in \Rb^{p \times d}$ and $\widehat{\Vv} = (\widehat{\vv}_1,\dots,\widehat{\vv}_d) \in \Rb^{p\times d}$ have orthonormal eigenvector columns satisfying $\Sigma \vv_j =\lambda_j \vv_j$ and $\widehat{\Sigma} \widehat{\vv}_j = \widehat{\lambda}_j \widehat{\vv}_j$ for $j=1,\dots,d$. Let $\Vc$ and $\widehat{\Vc}$ be the subspaces in $\Rb^p$ spanned by the columns of $\Vv$ and $\widehat{\Vv}$, respectively. Let $\Theta \in [0,\frac{\pi}{2}]$ be the largest principal angle between $\Vc$ and $\widehat{\Vc}$. Then,
    \begin{align*}
        \sin \Theta \leq \frac{2 \|\widehat{\Sigma} - \Sigma\|}{\lambda_d -\lambda_{d+1}}.
    \end{align*}
\end{lem}

Throughout, we write $\mbox{EC}_d(\mu, \Sigma)$ in place of $\mbox{EC}_d(\mu, \Sigma, \omega)$ for simplicity.

\subsubsection*{Proof of Proposition \ref{prop-privKg}} \label{pf-prop2.1}
\begin{proof}
First, note that $ \Delta_F(\widehat K_g) = \Delta_2(\vecd \circ \widehat K_g) $ since $ \|A\|_F = \|\vecd(A)\|_2 $.
Consider $ S = (x_1, \dots, x_n) $ and its neighbor $ S' = (x_1, \dots, x_n') $.
Then 
\begin{align*}
    \|\widehat K_g(S) - \widehat K_g(S')\|_F
    &= \frac{2}{n(n-1)} \sum_{i=1}^{n-1} \|g(x_n - x_i)g(x_n - x_i)^\top - g(x_n' - x_i)g(x_n' - x_i)^\top\|_F \\
    &\le \frac{2}{n(n-1)} \cdot (n-1) \cdot 2\|g\|_\infty^2 \\ 
    &= \frac{4\|g\|_\infty^2}{n}.
\end{align*}
This shows that $\Delta_2(\vecd \circ \widehat K_g) = \Delta_F(\widehat K_g) \le 4\|g\|_{\infty}^2/n $.
Then Proposition \ref{prop:gaussian-mech} implies that
$ (\vecd \circ \widehat K_g)(S) + \xi $ satisfies $ (\varepsilon, \delta) $-DP.
Finally, by taking $ \vecd^{-1} $, we get the desired result thanks to the post-processing property.
\end{proof}

\subsubsection*{Proof of Proposition \ref{prop:gkendall}}
\begin{proof}
Note that $(X - \widetilde X)/\sqrt{2} \sim \mbox{EC}_d(0, \Sigma)$. Therefore, Proposition 1 of \cite{raymaekers2019generalized} implies the desired result.
\end{proof}

\subsubsection*{Proof of Theorem \ref{thm:utility}}
\begin{proof}
    For $ 1 \le i < j \le d $, denote $ \kappa(i, j) := g(X_i - X_j)g(X_i - X_j)^\top $.
    By using the triangle inequality and boundedness of $g$, some straightforward computations imply that
    \begin{equation*}
        \left\|\kappa(i, j) - \Eb (\kappa(i, j))\right\|_2 \le \|g\|_{\infty}^2 + \|K_g\|_2,
    \end{equation*}
    and 
    \begin{equation*}
        \left\Vert \Eb \left\{\kappa(i, j) - \Eb (\kappa(i, j))\right\}^2 \right\Vert_2
        \le \|K_g\|_2(\|g\|_{\infty}^2 + \|K_g\|_2).
    \end{equation*}
    From the above two bounds, the matrix Bernstein's inequality for $U$-statistics 
    (Theorem B.2 of \cite{han2018eca}) implies that
    \begin{equation} \label{eq:bernstein}
    \Pb\left(\|\widehat K_g(S) - K_g\|_{2} \ge t\right)
    \le 
    d \exp \left(- \frac{3nt^2}{16\|K_g\|_2(\|g\|_{\infty}^2 + \|K_g\|_2)}\right)
    \end{equation}
    for any $ 0 \le t \le \|K_g\|_2  $.
    To make the left-hand side of \eqref{eq:bernstein} to be bounded by $ \beta / 2 $, $ t $
    should be greater than 
    \begin{equation*}
        t \ge \|K_g\|_2 \sqrt{\frac{16(\|g\|_\infty^2/\|K_g\|_2 + 1)\log(2d/\beta)}{3n}}.
    \end{equation*}
    Also, to utilize the bound \eqref{eq:bernstein}, $ t $ should be less than or equal to $ \|K_g\|_2 $.
    For this, we assume a sufficiently large sample size as 
    $ n \ge \frac{16}{3} (\|g\|_\infty^2/\|K_g\|_2 + 1)\log(2d/\beta)$.
    Then, we have 
    \begin{equation}\label{eq:pf-thm8-1}
        \Pb \left( \|\widehat K_g(S) - K_g\|_2 \le 
        \|K_g\|_2 \sqrt{\frac{16(\|g\|_\infty^2/\|K_g\|_2 + 1)\log(2d/\beta)}{3n}}
        \right) \le \beta/2.
    \end{equation}

    Next, we give a high-probability bound for 
    $ \|\widetilde K_g(S) - \widehat K_g(S)\|_2 = \|\vecd^{-1}(\xi)\|_2 $, where
    $ \xi_i \overset{i.i.d.}{\sim} N(0, \sigma_{\varepsilon, \delta}) $.    
    Suppose that $ t \ge \sqrt{d} .$ 
    According to the concentraion inequality for sub-Gaussian entried symmetric matrix 
    (Corollary 4.4.8 of \cite{vershynin2018high}),
    we have 
    \begin{equation} \label{eq:pf-thm8-2}
        \Pb \left( \|\vecd^{-1}(\xi)\|_2 
            \le C \sigma_{\varepsilon, \delta} \left(\sqrt{d} + \sqrt{\log(8/\beta)}\right)
        \right) \le \beta/2,
    \end{equation}
    where $ C > 0 $ is a generic absolute constant.
    
    On the other hand, the variant of Davis-Kahan $ \sin \Theta $ theorem (Lemma \ref{lem:kahan})
    implies that 
    \begin{equation} \label{eq:pf-thm8-3}
        \sin \Theta \left(\col(\widetilde V_m^g(S)), \col(V_m)\right) \le
        \frac{2\left(
            \|\widetilde K_g(S) - \widehat K_g(S)\|_2 + \|\widehat K_g(S) - K_g\|_2\right)}
            {\phi_{g, m} - \phi_{g, m+1}}.
    \end{equation}
    Then, put \eqref{eq:pf-thm8-1} and \eqref{eq:pf-thm8-2} into \eqref{eq:pf-thm8-3}, and by substituting 
    \[
        \sigma_{\varepsilon, \delta} =  \frac{4\|g\|_{\infty}^2\sqrt{2 \log(1.25/\delta)}}{n \varepsilon},
    \]
    we can obtain the $ 1-\beta $ probability upper bound as in the theorem.
    This completes the proof.
\end{proof}

\subsubsection*{Proof of Theorem \ref{thm:bp-Kendalltau}}

\begin{proof}
    Let $ \alpha \in [0, 1) $. 
    For a given dataset $ S = \{x_1, \dots, x_n\} $, consider $ \alpha $-corrupted dataset     
    \[
        S_{\alpha} = \{x_1, \dots, x_{(1-\alpha)n}, y_1, \dots, y_{n \alpha}\}.
    \]
    Then, by the triangle inequality, we have
    \begin{align*}
        &\|\widehat K_g(S_{\alpha}) - \widehat K_g(S)\| \\
        &= \frac{1}{\binom{n}{2}} 
        \bigg\| \sum_{1 \le i \le (1-\alpha)n} \sum_{1 \le j \le n \alpha} 
        g(x_i - y_j)g(x_i - y_j)^\top -  g(x_i - x_j)g(x_i - x_j)^\top  \\
        &\quad + \sum_{1 \le i < j \le n \alpha} g(y_i - y_j)g(y_i - y_j)^\top -  
        g(x_i - x_j)g(x_i - x_j)^\top
        \bigg\| \\
        &\le \frac{1}{\binom{n}{2}} \cdot 2 \|g\|_{\infty}^2 \cdot
        \left((1-\alpha)n \cdot \alpha n + \binom{n \alpha}{2} \right)  \\
        &\le 4 \alpha \|g\|_{\infty}^2.
    \end{align*}
    Hence, the variant of Davis-Kahan $ \sin \Theta $ theorem (Lemma \ref{lem:kahan})
    implies that 
    \begin{equation} \label{eq:thm6pf-eq1}
        \sin \Theta (\Vc_m^g(S_{\alpha}), \Vc_m^g(S))
        \le \frac{2 \|\widehat K_g(S_{\alpha}) - \widehat K_g(S)\|}{\hat \phi_{g, m} - \hat \phi_{g, m+1}}
        \le \frac{8 \alpha \|g\|_{\infty}^2}{\hat \phi_{g, m} - \hat \phi_{g, m+1}}.
    \end{equation}
    Now, suppose that $\textnormal{\mbox{bp}}(\mathcal{V}_{m}^g; S) = \alpha_0$.
    Then, from the definition, it holds that for any $ \gamma \in (0, 1) $, there exists 
    a $ \alpha_0 $-corrupted dataset $ S_{\alpha_0} $ satisfies
    \begin{equation} \label{eq:thm6pf-eq2}
        1 - \gamma \le \sin \Theta (\Vc_m^g(S_{\alpha_0}), \Vc_m^g(S)).
    \end{equation}
    By combining \eqref{eq:thm6pf-eq1} and \eqref{eq:thm6pf-eq2}, we get
    \begin{equation*}
        1 - \gamma \le \frac{8 \alpha_0 \|g\|_{\infty}^2}{\hat \phi_{g, m} - \hat \phi_{g, m+1}},
    \end{equation*}
    and hence 
    \begin{equation*}
        \alpha_0 \ge \frac{\hat \phi_{g, m} - \hat \phi_{g, m+1}}{8 \|g\|_{\infty}^2}.
    \end{equation*}
    This completes the proof.
\end{proof}

\subsubsection*{Proof of Theorem \ref{thm:corrupted}}
\begin{proof}
    As in the proof of Theorem \ref{thm:utility}, 
    the variant of Davis-Kahan $ \sin \Theta $ theorem (Lemma \ref{lem:kahan})
    implies that 
    \[
        \sin \Theta \left(\col(\widetilde V_m^g(S_{\alpha})), \col(V_m)\right) \lesssim
        \frac{\left(
        \|\vecd^{-1}(\xi)\|_2 + \|\widehat K_g(S_{\alpha}) - K_g\|_2\right)}
        {\phi_{g, m} - \phi_{g, m+1}}.
    \]
    We already know an upper bound of $ \|\vecd^{-1}(\xi)\|_2 $ described in \eqref{eq:pf-thm8-2}.
    This yields a bound $ \|\vecd^{-1}(\xi)\|_2 \lesssim \|g\|_\infty^2 I_1 $ with probability at least
    $ 1 - \beta/2 $.

    Next, it is enough to derive an upper bound of $ \|\widehat K_g(S_{\alpha}) - K_g\|_2 $.
    For this, observe that 
    \begin{align*}
        \left(\widehat K_g(S_{\alpha}) - K_g \right) 
        &= 
        \binom{n}{2}^{-1}
        \bigg( \sum_{1 \le i < j \le (1-\alpha)n}  
        \left(g(X_i - X_j)g(X_i - X_j)^\top -  K_g\right) \\
        &\quad + \sum_{1 \le i \le (1-\alpha)n} \sum_{1 \le j \le n \alpha} 
        \left(g(X_i - y_j)g(X_i - y_j)^\top -  K_g\right)  \\
        &\quad + \sum_{1 \le i < j \le n \alpha} \left(g(y_i - y_j)g(y_i - y_j)^\top - K_g\right)
        \bigg) \\
        &=: L_1 + L_2 + L_3.
    \end{align*}

    To bound $ \|L_1\| $ one can adapt the bound \eqref{eq:pf-thm8-1} to uncorrupted data of
    $ X_1, \dots, X_{(1-\alpha) n} $.
    This leads to the bound that $ \|L_1\| \le \|g\|_\infty^2 I_2 $ with probability at least
    $ 1 - \beta/2 $.
    
    For the $ L_2 + L_3 $, we use boundedness of $ g $. 
    Simply,
    \begin{align*}
        \|L_2 + L_3\|
        &\le \frac{1}{\binom{n}{2}} \cdot (\|g\|_{\infty}^2 + \|K_g\|_2) \cdot
        \left((1-\alpha)n \cdot \alpha n + \binom{n \alpha}{2} \right)  \\
        &\lesssim \|g\|_{\infty}^2 I_3.
    \end{align*}
    By gathering all these bounds, we prove the theorem.
\end{proof}

\subsubsection*{Proof of Equation \eqref{eq:eigval-wins}}
We use the following lemma which was intrinsically proved in the proof of Proposition 1 of 
\cite{raymaekers2019generalized}.

\begin{lem}{B.1} \label{lem:eig-gsscm}
    Suppose that $X \sim \mbox{EC}_d(0, \Sigma)$
    and denote the eigendecomposition of $\Sigma$ as $\Sigma = U \Lambda U^\top$, where $ U \in \Oc(d) $ and
    $ \Lambda = \diag(\lambda_1, \dots, \lambda_d) $ with $ \lambda_1 \ge \dots \ge \lambda_d > 0 $.
    Then, the $i$th eigenvalue of $\Eb\left[g(X)g(X)^\top\right]$, denoted $\lambda_{g, i}$, is expressed as
    \begin{equation} 
        \lambda_{g, i} 
        = \Eb_Z\left[ ([g(\sqrt{\Lambda} Z)]_i)^2\right].
    \end{equation}
\end{lem}

\begin{proof}[Proof of \eqref{eq:eigval-wins}]
Let $X$ be an elliptically distributed random vector, and let $\widetilde X$ be an independent copy of $X$. Define $W = \frac{X - \widetilde X}{\sqrt{2}}$. 
Then $ W $ becomes a centered elliptical distribution with dispersion matrix $ \Sigma $.
Also, $ W $ has a stochastic representation of $ W = \Sigma^{1/2} Z' $ for a spherically 
symmetric distribution $ Z' $.
Since $ Z' $ is spherically symmetric, it can further be written as 
$ Z' \overset{d}{=} R\textbf{S} $ where $ \textbf{S} \sim \mbox{Unif}(\Sb^{d-1})$ and $ R \ge 0 $ is a positive random variable independent to $ \textbf{S} $.
Note that $ R^2 \overset{d}{=} \|Z'\|_2^2 \overset{d}{=} \|\Sigma^{-1/2}W\|_2^2 $.

By applying Lemma \ref{lem:eig-gsscm} to $ W $, we get
\[
    \phi_{w, \ell}^{(r)} 
    = \Eb_{Z'} \left[ \left(g_{wins}^{(r)}(\sqrt{\Lambda} Z')_\ell\right)^2 \right].
\]
Observe that 
\[
g_{wins}^{(r)}(\sqrt{\Lambda} Z')_\ell = 
(\sqrt{\Lambda}Z')_i I(\|\sqrt{\Lambda}Z'\|_2 \le r) + r \cdot \frac{(\sqrt{\Lambda}Z')_i}{\|\sqrt{\Lambda}Z'\|_2} I(\|\sqrt{\Lambda}Z'\|_2 > r).
\]
Denote $ a \wedge b := \min(a, b) $.
By substituting $ Z' = R\mathbf{S} = (RS_1, \dots, RS_d)^{\top} $, we have
\begin{align*}
    \left(g_{wins}^{(r)}(\sqrt{\Lambda} Z')_\ell\right)^2
    &\overset{d}{=}  R^2 \lambda_\ell S_\ell^2 I(R^2 \|\sqrt{\Lambda}S\|_2^2 \le r^2 )
    + \frac{r^2}{R^2 \|\sqrt{\Lambda}S\|_2^2} \lambda_\ell R^2 S_\ell^2 
    I(R^2 \|\sqrt{\Lambda}S\|_2^2 > r^2) \\
    &=R^2 \cdot \lambda_\ell S_\ell^2 I\left(R^2 \le \frac{r^2}{\|\sqrt{\Lambda}S\|_2^2}\right) 
    + \frac{r^2}{\|\sqrt{\Lambda}S\|_2^2} \cdot \lambda_\ell S_\ell^2 
    I\left(R^2 > \frac{r^2}{\|\sqrt{\Lambda}S\|_2^2}\right) \\
    &= \left(R^2 \wedge \frac{r^2}{\|\sqrt{\Lambda}S\|_2^2} \right) \lambda_\ell S_\ell^2 \\
    &= \left(R^2 \wedge \frac{r^2}{\sum_j \lambda_j S_j^2} \right)\lambda_\ell S_\ell^2.
\end{align*}

By summing up, we finally have
    \[
        \phi_{w, \ell}^{(r)} 
        = \Eb\left[\left(R^2 \wedge \frac{r^2}{\sum_j \lambda_j S_j^2} \right)\lambda_\ell S_\ell^2 \right],
\]
    where $R^2 \overset{d}{=} \|\Sigma^{-1/2}W\|_2^2 
    = \frac{1}{2}(X - \widetilde X)^\top \Sigma^{-1}(X - \widetilde X) $ and 
    $(S_1, \dots, S_d) \sim \mbox{Unif}(\Sb^{d-1})$.
This completes the proof.
\end{proof}

\subsubsection*{Proof of Proposition \ref{prop:errbound-sph}}
\begin{proof}
    From Theorem 3.2 of \cite{han2018eca}, we have $\lambda_j/\tr(\Sigma) \asymp \phi_{sph, j}$ as $d$ increases.
    Then 
    \[
    \frac{\phi_{sph, 1}}{\phi_{sph, m} - \phi_{sph, m+1}} \asymp \frac{\lambda_1/\tr(\Sigma)}{\lambda_m/\tr(\Sigma) - \lambda_{m+1}/\tr(\Sigma)} = \frac{\lambda_1}{\lambda_m - \lambda_{m+1}}
    \]
    and
    $\|g_{sph}\|_\infty^2 / \phi_{g, 1} \asymp \tr(\Sigma) / \lambda_1 = \er(\Sigma)$.
    Plugging this into $u_{g, m}$ gives the desired result.
\end{proof}

\subsubsection*{Proof of Proposition \ref{prop:errbound-wins}}
\begin{proof}
    Observe that
    \[
        \phi_{w, \ell}^{(r)}  \le 
        \Eb\left[R^2 \lambda_\ell S_\ell^2\right] \wedge
        \Eb\left[\frac{r^2}{\sum_{j=1}^d \lambda_j S_j^2} \lambda_\ell S_\ell^2\right]
        \le r^2 \phi_{sph, \ell}.
    \]
    For the first inequality, we use the fact that $\min(a, b) = \frac{a+b - |a-b|}{2}$; for any $a, b \in \Rb$,
    \[
    \Eb(a \wedge b) 
    = \frac{\Eb a + \Eb b - \Eb|a - b|}{2}
    \le \frac{\Eb a + \Eb b - |\Eb a - \Eb b|}{2}
    = \Eb a \wedge \Eb b.
    \]

    To achieve a lower bound we do conditioning as follows.
    \begin{align*}
        \phi_{w, \ell}^{(r)} 
        &= \Eb\left[ \Eb \left[ \min \left(R^2, \frac{r^2}{\sum_j \lambda_j S_j^2} \right)
        \lambda_\ell S_\ell^2 \:\bigg\vert\: 
            (S_1, \dots, S_d)
        \right] \right] \\
        &\ge \Eb \left[
            \Pb\left( R^2 \ge \frac{r^2}{\sum_j \lambda_j S_j^2 } \:\bigg\vert\: 
            (S_1, \dots, S_d) \right) \frac{r^2 \lambda_\ell S_\ell^2}{\sum_j \lambda_j S_j^2}
        \right] \\
        &\ge \Eb \left[
            \Pb\left( R^2 \ge \frac{r^2}{\lambda_d} \:\bigg\vert\: 
            (S_1, \dots, S_d) \right) \frac{r^2 \lambda_\ell S_\ell^2}{\sum_j \lambda_j S_j^2}
        \right] \\
        &= \Pb\left(R^2 \ge \frac{r^2}{\lambda_d}\right) \cdot r^2 \phi_{sph, \ell}.
    \end{align*}
    This completes the proof of the first part.

    For the second part, suppose that $ \phi_{wins, \ell}^{(r)} \asymp r^2 \phi_{sph, \ell} $.
    Observe that 
    \[
        \frac{\phi_{w, 1}^{(r)}}{\phi_{w, m}^{(r)} - \phi_{w, m+1}^{(r)}}
        \asymp 
        \frac{r^2\phi_{sph, 1}}{r^2\phi_{sph, m} - r^2\phi_{wins, m+1}}
        \asymp 
        \frac{\phi_{sph, 1}}{\phi_{sph, m} - \phi_{wins, m+1}}
    \]
    and
    \[
        \frac{\|g_{wins}^{(r)}\|_\infty^2}{\phi_{wins, 1}^{(r)}}
        \asymp 
        \frac{r^2}{r^2 \phi_{sph, 1}}
        = \frac{\|g_{sph}\|_\infty^2}{\phi_{sph, 1}}.
    \]
    Then, from the expression of $u_{g, m}$, it can be easily checked that 
    \[
        u_{wins, m}^{(r)} \asymp u_{sph, m}.
    \]
    This completes the proof.
\end{proof}

\section{Additional numerical studies} \label{app-add_num}

\renewcommand{\thefigure}{B.\arabic{figure}}
\setcounter{figure}{0}

\subsection{Simulation study comparing GSSCM and \texorpdfstring{$g$}-DPPCA} \label{app-rmk2-2sim}
In this section, we provide simulation results comparing the performance of GSSCM with private mean estimation and $g$-DPPCA methods, as discussed in Remark \ref{remark_2-2}.

Recall that GSSCM is defined as 
$$\Sigma_{g} = \Eb\left[ g(X - \mu)g(X - \mu)^\top \right],$$ 
where $\mu$ is a location parameter.
To construct private PC estimator based on $\Sigma_g$, we need to estimate $\mu$ in a privacy-preserving way. 
For this, we use the private modified winsorized mean (PMWM) proposed by \cite{ramsay2025improved}.
Since PMWM is better suited with the use of the zero-concentrated differential privacy (zCDP) \citep{bun2016concentrated}, 
the detailed procedure of PMWM, given below, accounts for the privacy loss by the $\rho$-zCDP. Here, $\rho > 0$ is a given privacy budget. 
\begin{enumerate}
    \item Let $S = (X_1, \dots, X_n)$ be an observed random sample. 

    \item Calculate PMWM from $S$ using the budget $\rho/2$, and denote it as $\tilde \mu$.

    \item Get $ \widehat{\Sigma}_g(S; \tilde \mu) = \frac{1}{n}\sum_{i=1}^{n}g(X_i - \tilde \mu)g(X_i - \tilde \mu)^\top $.

    \item Apply $\frac{\rho}{2}$-zCDP additive Gaussian mechanism to $\widehat{\Sigma}_g$: Get $ \widetilde{\Sigma}_g = \widehat{\Sigma}_g(S; \tilde \mu) + \vecd^{-1}(\xi) $, where $\xi \sim N_{d(d+1)/2}(0, \sigma^2 I_d)$ and $ \sigma = 2\|g\|_{\infty}^2 / (n\sqrt{4\rho})$.
    
    \item Finally, get $ \widetilde{V}_g $ as the eigenvector matrix of $ \widetilde{\Sigma}_g $.
\end{enumerate}

For a fair comparison, the Gaussian noise in our $g$-DPPCA proposal is also calibrated to satisfy $\rho$-zCDP.  

For the numerical comparisons, we consider same data setting of Gaussian and multivariate $ t_1 $ with two-spiked covariance as in Section \ref{sec:sim-study}.
We conduct simulation under parameters of 
$ n \in \{250, 500, 1000, 1500\} $, $ d \in \{10, 25\} $ 
and $ \rho \in \{0.1, 0.5\} $.
For each generated datset, we fit $g$-DPPCA and GSSCM with PMWM as explained, where $g_{sph}$ and $ g_{wins}^{(r)} $ with $ r = \sqrt{d} $ are considered.
Estimation loss is measured by $ \|\widetilde{V}_2\widetilde{V}_2^\top - V_2 V_2^\top\|_F $ and each simulation is repeated 25 times.

\begin{figure}[tbh]
    \centering
    \begin{subfigure}{0.48\textwidth}
        \centering
        \includegraphics[width=\linewidth]{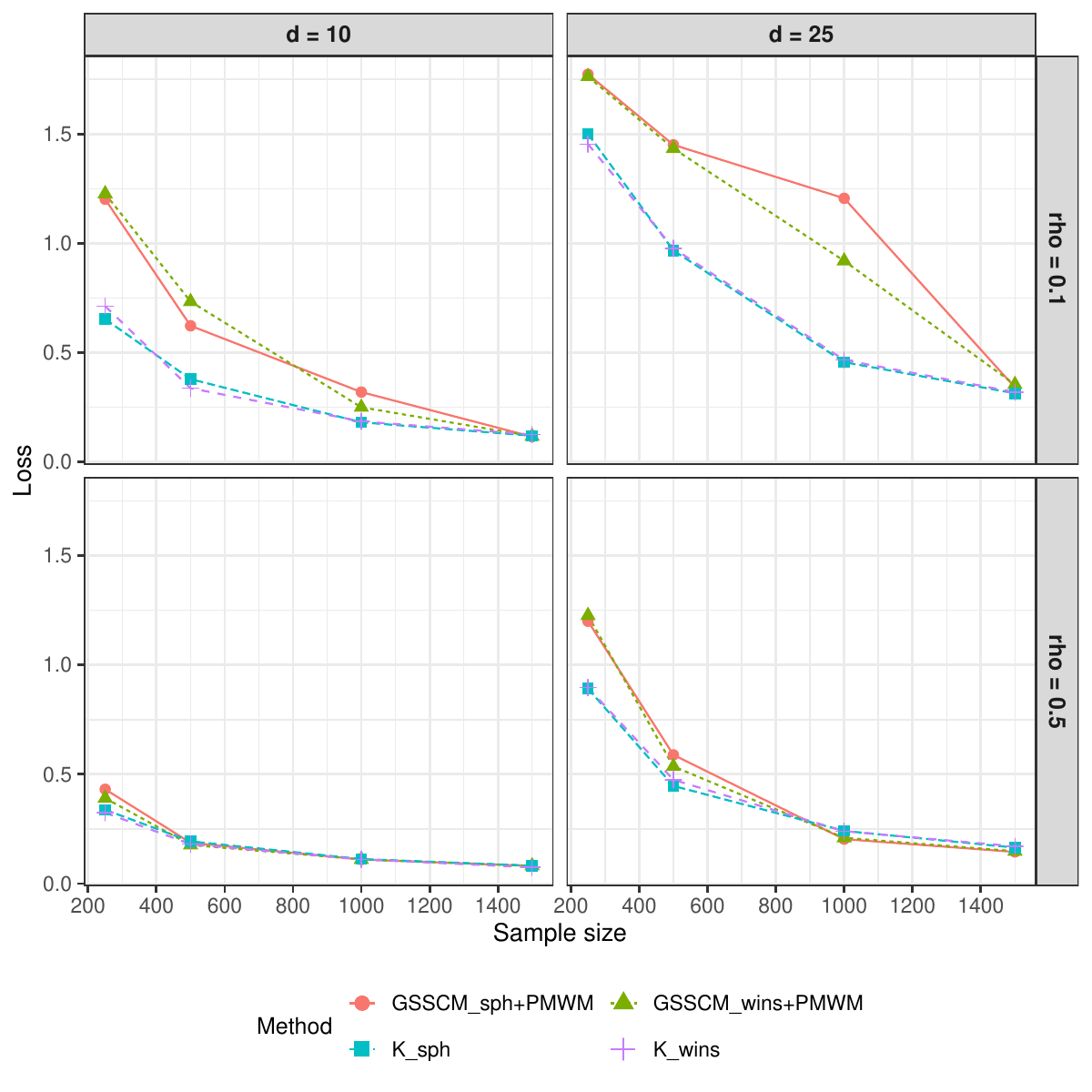}
        \caption{$ X \sim N(0, \Sigma) $}
        \label{fig:GSSCM-PMWM-sph}
    \end{subfigure}
          \hfill
    \begin{subfigure}{0.48\textwidth}
        \centering
        \includegraphics[width=\linewidth]{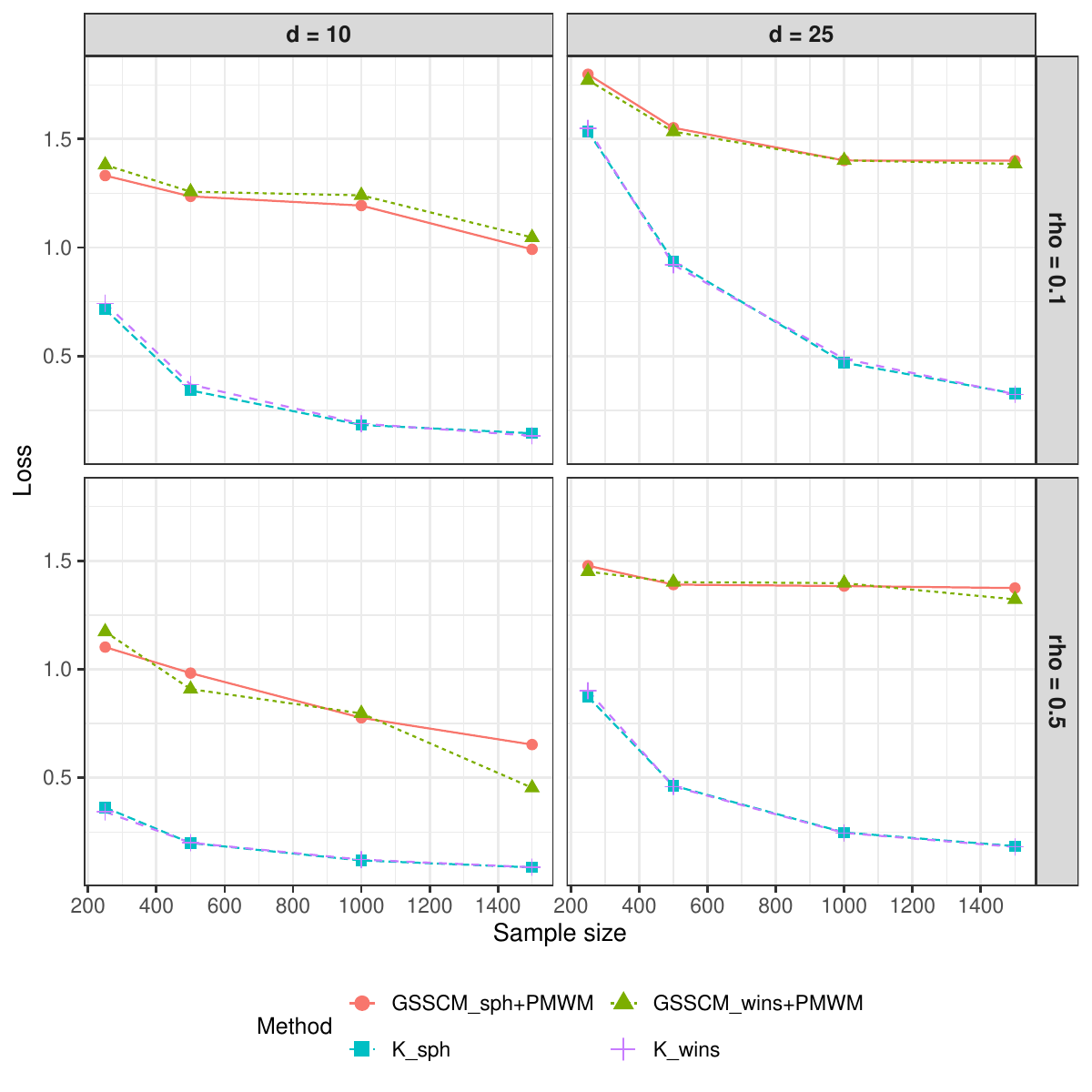}
        \caption{$X \sim t_1(0, \Sigma)$}
        \label{fig:GSSCM-PMWM-wins}
    \end{subfigure}
    \caption{
        Averaged losses over different settings.
    }
    \label{fig:GSSCM-PMWM-appendix}
\end{figure}

The simulation results are reported in Figure \ref{fig:GSSCM-PMWM-appendix}.
In all cases, $g$-DPPCA based PCA methods outperform GSSCM--PMWM methods.
In Gaussian case of (a), GSSCM--PMWM show similar performance compared to $g$-DPPCA.
However, the performance gap between GSSCM--PMWM and $ g $-DPPCA does not narrow even if the sample size increases when data follows the multivariate $t_1$ distribution as in (b).
Interestingly, the performance gap is small when the dimension is low, the privacy requirement is weak, and the data follow a light-tailed distribution.
This can be attributed to two main reasons: (i) the privacy budget is split between mean estimation and the additive mechanism, and (ii) errors in mean estimation cause a failure to properly center the data before the transformation. In particular, errors in mean estimation are typically large for high-dimension, high-privacy or heavy-tailed distributions.
Our proposed methods, based on Kendall's tau, are free from these issues and therefore perform better than the GSSCM–PMWM methods.

\subsection{Simulation study on the choice of winsorization radius} \label{app-sim-radius}
In this subsection, we discuss how $g_{wins}$-DPPCA is affected by the choice of winsorization radius $r$, through  numerical simulations. In short, we find that the performance remains similar across different radii, as long as the radius lies within $[1,\sqrt{d}]$ or is chosen as a quantile of the observed paired differences.

\begin{figure}[]
    \centering
    \begin{subfigure}{0.75\textwidth}
        \centering
        \includegraphics[width=\linewidth]{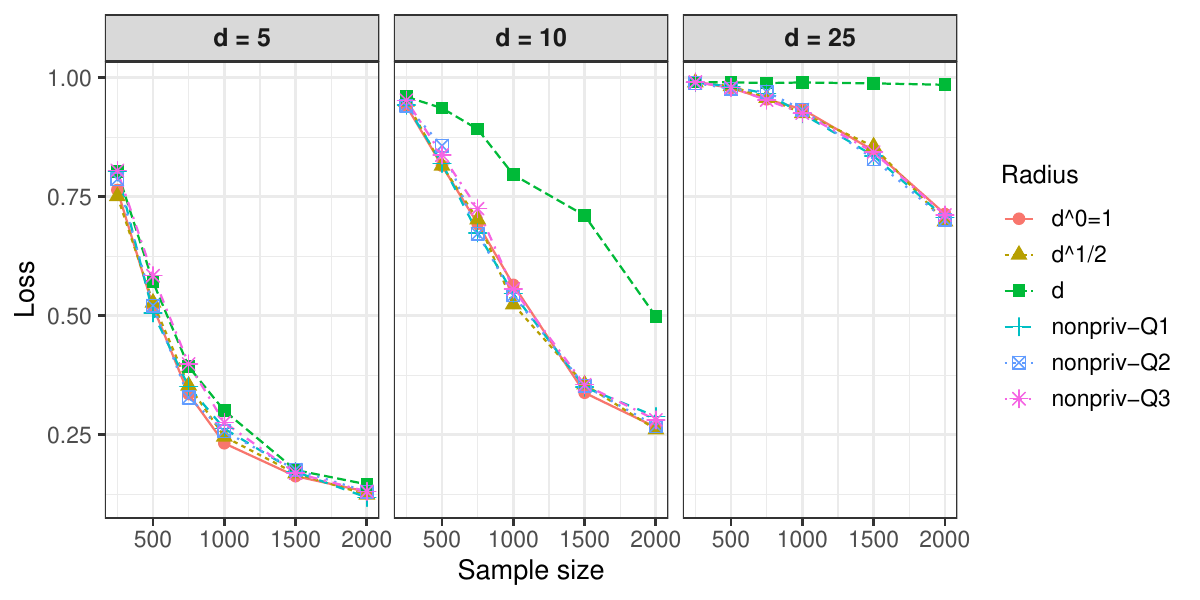}
        \caption{Gaussian distribution}
    \end{subfigure}
          \vspace{5pt}
          \\
    \begin{subfigure}{0.75\textwidth}
        \centering
        \includegraphics[width=\linewidth]{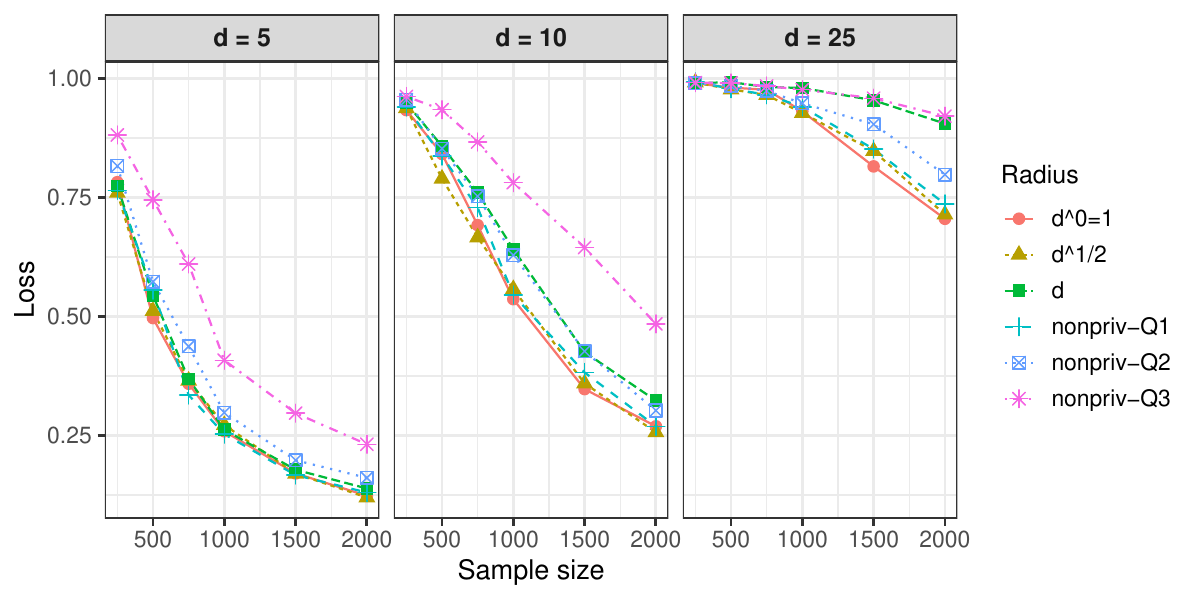}
        \caption{$t_1$ distribution}
    \end{subfigure}
    \caption{
        The effect of winsorization radius (non-private quantiles)
    }
    \label{fig:winsor_radius_nonprivquantile_sim}
    \end{figure}

\begin{figure}[]
    \centering
    \begin{subfigure}{0.75\textwidth}
        \centering
        \includegraphics[width=\linewidth]{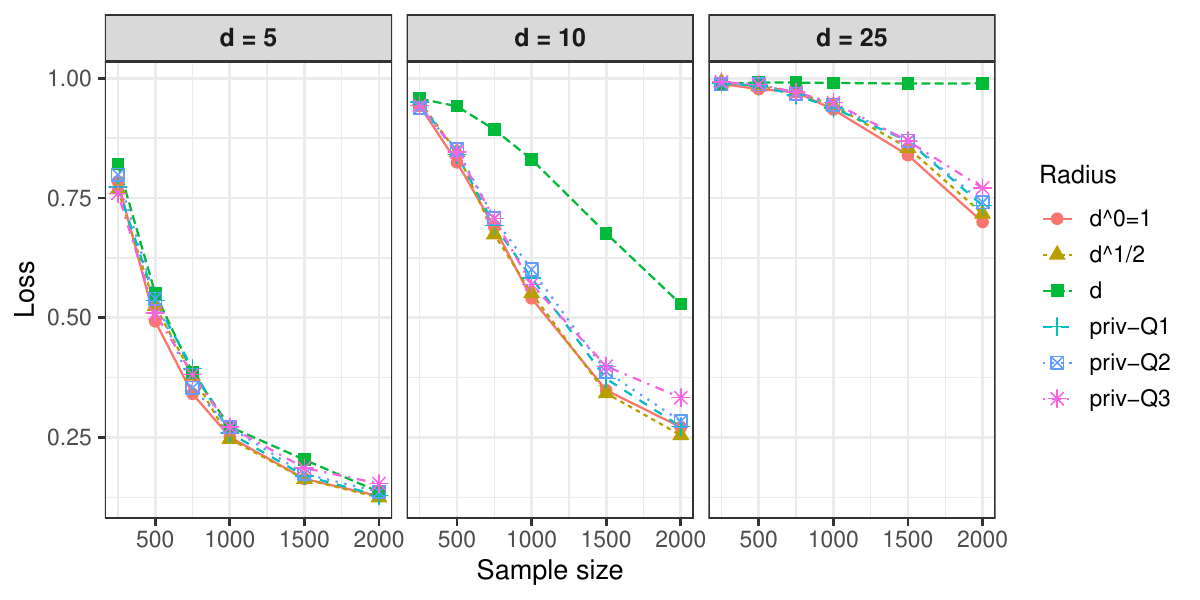}
        \caption{Gaussian distribution}
    \end{subfigure}
          \\
          \vspace{5pt}
    \begin{subfigure}{0.75\textwidth}
        \centering
        \includegraphics[width=\linewidth]{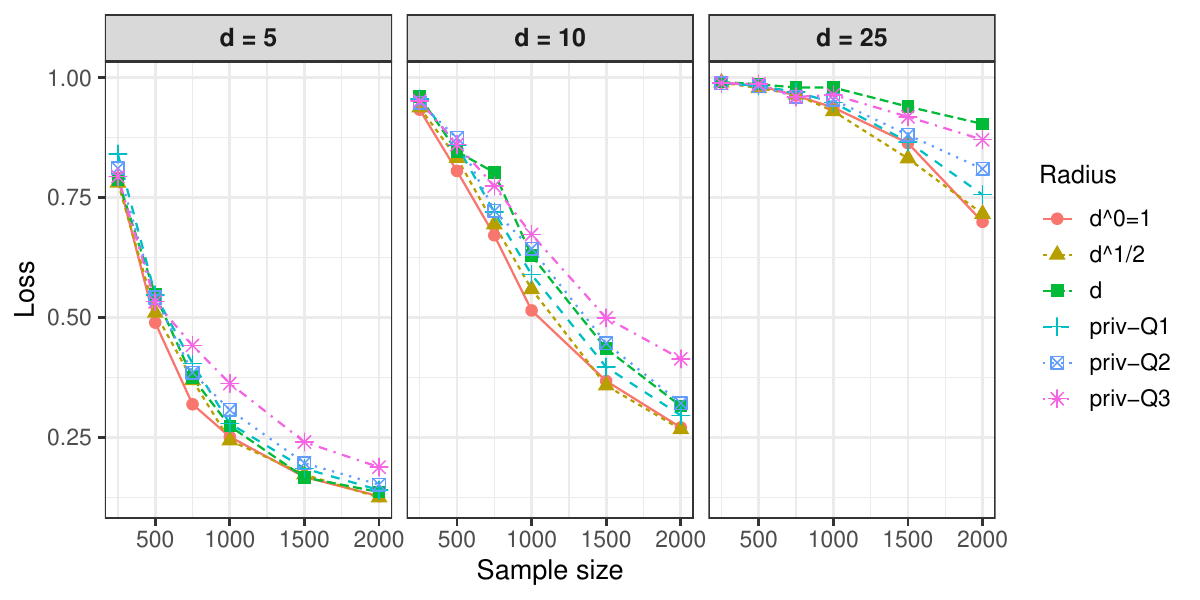}
        \caption{$t_1$ distribution}
    \end{subfigure}
    \caption{        
        The effect of winsorization radius (private quantiles)
    }
    \label{fig:winsor_radius_sim}
    \end{figure}

\begin{figure}[]
    \centering
    \begin{subfigure}{0.75\textwidth}
        \centering
        \includegraphics[width=\linewidth]{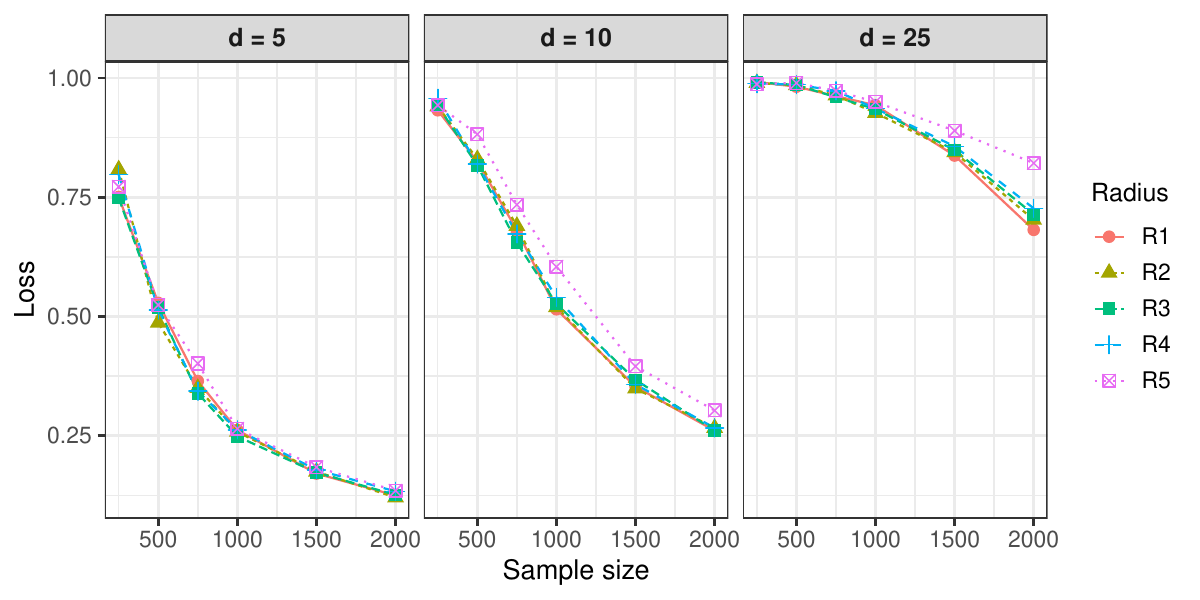}
        \caption{Gaussian distribution}
    \end{subfigure}
    \\
    \begin{subfigure}{0.75\textwidth}
        \centering
        \includegraphics[width=\linewidth]{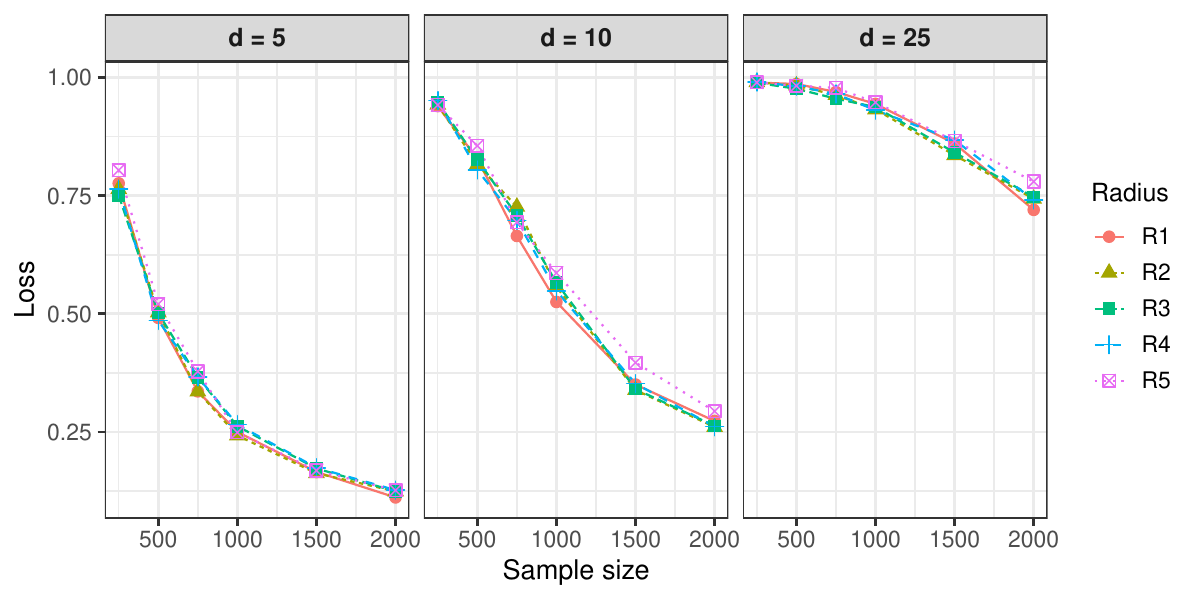}
        \caption{$t_1$ distribution}
    \end{subfigure}
    \caption{
        The effect of winsorization radius $r = c\sqrt{d}$ for $c \in \{0.25, 0.5, 1, 1.5, 2.0\}$. The legend in the figure indicates $R_i = c\sqrt{d}$ so that $R_1 < R_2 < R_3 < R_4 < R_5. $ 
    }
    \label{fig:winsor_radius_sens_sim}
    \end{figure}

We first examine the appropriate scale of the winsorization radius. For this, we compare the effects of different radii $r \in {1, \sqrt{d}, d}$ and find that choosing $r \asymp d$ is typically too large, resulting in the loss of robustness of the generalized Kendall’s tau and excessively large calibrated noise. This pattern can be observed in Figures~\ref{fig:winsor_radius_nonprivquantile_sim} and~\ref{fig:winsor_radius_sim}. In contrast, setting the radius on the order of $\sqrt{d}$ yields the best performance. Moreover, as long as $r \asymp \sqrt{d}$, the specific coefficient $c$ in $r = c\sqrt{d}$ has little effect on performance. To verify this, we considered $c \in {0.25, 0.5, 1, 1.5, 2.0}$ and compared the results. As shown in Figure~\ref{fig:winsor_radius_sens_sim}, the choice of $c$ leads to only minor performance differences except for $c = 2.0$. These results suggest that setting $r \propto \sqrt{d}$ provides a sufficiently good choice of radius.

Second, the winsorization radius can be successfully determined from the observed data. In our theoretical analysis, we used $r = \sqrt{d\lambda_d}$ as a reference choice. We emphasize that it is possible to select an appropriate winsorization radius solely from the observed data, without knowing any parameter of the underlying distribution. A natural approach is to set the radius as a quantile of the paired differences ${|X_i - X_j|_2/\sqrt{2}}$. 
The performances of $g_{wins}$-DPPCA, when the radius is chosen as the non-private Q1, or Q2, are similar to the performance of choosing $r = 1$ (see Figure \ref{fig:winsor_radius_nonprivquantile_sim}). Choosing Q3 appears to be too large for heavy-tailed distributions.

    However, to guarantee differential privacy, we should consider a private version of quantiles.
    For this, we propose to use differentially private quantile method proposed by \cite{durfee2024unbounded} to determine the radius.
    We split the total privacy budget $(\varepsilon, \delta)$ to $(0.05\varepsilon, 0)$ for quantile estimation and $(0.95\varepsilon, \delta)$ for $g_{wins}$-DPPCA.
    For a given $q \in \{0.25, 0.5, 0.75\}$, set a radius as the private $q$-quantile of $\{\|X_{i+n/2} - X_i\|_2/\sqrt{2}: 1 \le i \le n/2 \}$.
    Then conduct winsorized DPPCA with selected radius.   
    The corresponding simulation results are presented in Figure \ref{fig:winsor_radius_sim}.
    Although winsorized DPPCA with private quantiles result in a slight decreasing in performance, the overall difference compared to non-private quantiles is not significant. 
    

In summary, we recommend the following for a practice choice of $r$:
\begin{enumerate}
    \item[(i)] privately estimate the first quartile of the paired differences with low privacy budget as the value for $r$;
    \item[(ii)] use $r = c\sqrt{d}$ with together with prior information on $c$;
    \item[(iii)] Alternatively, simply use small $r$, effectively using the spherical transformation $g = g_{sph}$ in place of the winsorization.
\end{enumerate}

\subsection{Additional simulation on contaminated Gaussian model} \label{app-sim-contam}
In this subsection, we provide two additional simulation results and related discussions under the contaminated Gaussian model to examine robustness of proposed methods.
    
In the numerical studies of Section 4.2, recall that we considered a 2-spike covariance 
    \[
    \Sigma = (\lambda_1-\lambda_d)v_1v_1^\top + (\lambda_2-\lambda_d) v_2 v_2^\top + \lambda_dI_d
    \]
    and generated the contaminated Gaussian data from the following model:
    \begin{equation} \label{eq:r2q8-Xc}
    X^c \sim (1-\rho) N(0, \Sigma) + \rho N_d(c_{\perp}v_{\perp}, 0.05^2 I_d),
    \end{equation}
    where $c_{\perp} > 0$ and $v_{\perp} \in \Rb^d$ is an unit vector orthogonal to $v_1$ and $v_2$.
    In the original manuscript, we set $\rho = 0.05$ (i.e. 5\% of dataset is contaminated) and $c_{\perp} = 2.5\lambda_1$.
    The privacy parameter was fixed as $(\varepsilon, \delta) = (0.5, 10^{-5})$, and we remain those values in the following additional simulations.
    
    \begin{figure}[]
    \centering
    \begin{subfigure}{0.48\textwidth}
        \centering
        \includegraphics[width=\linewidth]{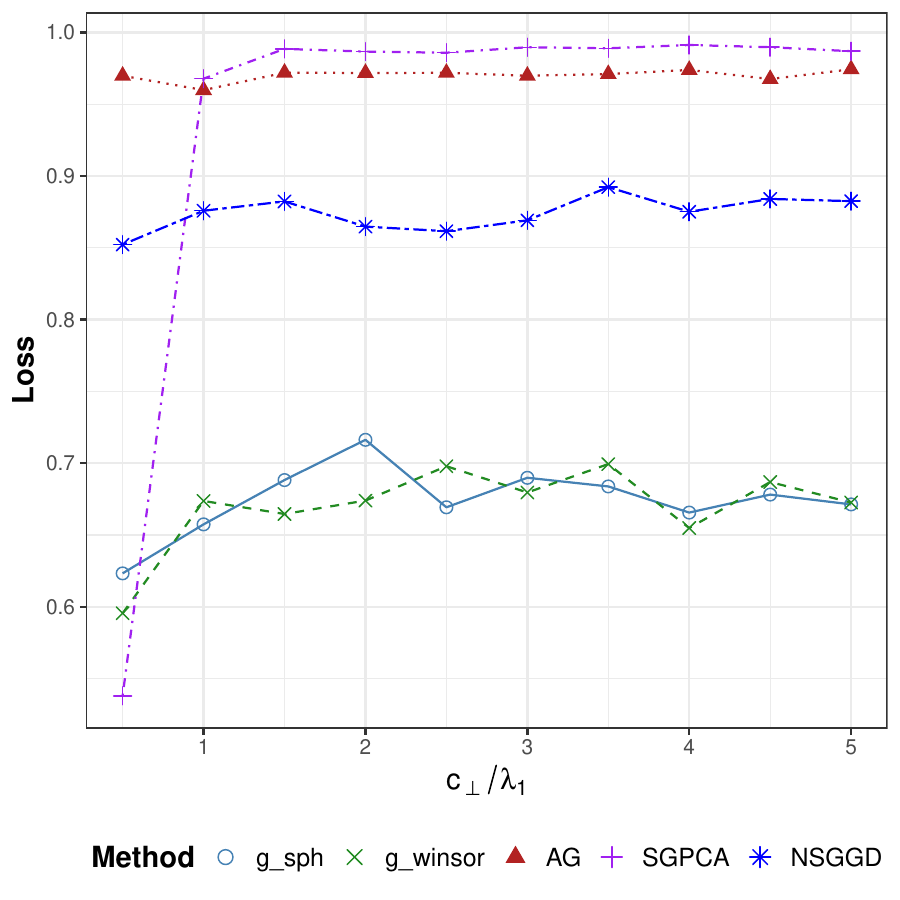}
        \caption{Varying $c_{\perp}/\lambda_1$.}
        \label{fig:sim_contam_cperp}
    \end{subfigure}
    \hfill
    \begin{subfigure}{0.48\textwidth}
        \centering
        \includegraphics[width=\linewidth]{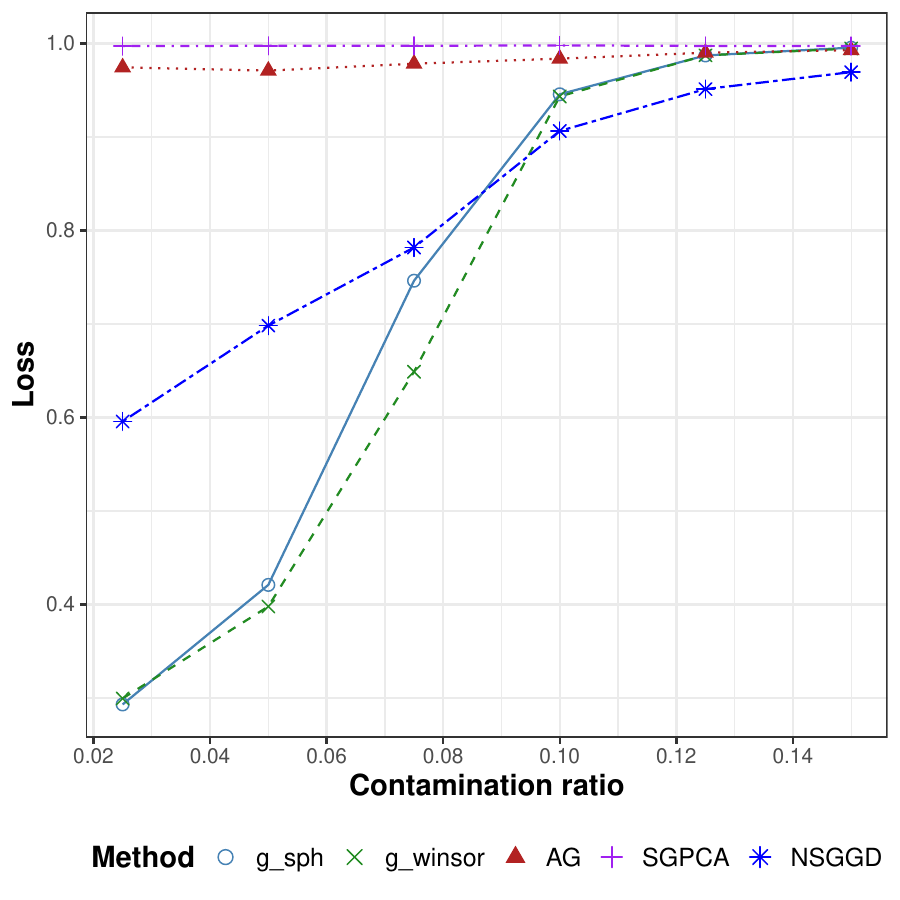}
        \caption{Varying $\rho$.}
        \label{fig:sim_contam_rho}
    \end{subfigure}
    \caption{
       Results of simulations comparing the performance of private PCA methods under the scenarios of (a) different magnitudes of contamination by varying $c_{\perp}/\lambda_1$ from 0.5 to 5 with fixed $(N, d, \rho) = (1000, 10, 5\%)$, and (b) different  contamination ratio $\rho$ from 2.5\% to 15\% with a fixed $(N, d, c_{\perp}) = (2000, 10, 2.5\lambda_1)$. 
    }
    \label{fig:sim_contam}
    \end{figure}

    First, we fit $g$-DPPCA and competing PCA methods to contaminated Gaussian data for different magnitudes of contamination by changing $c_{\perp}/\lambda_1$ fro 0.5 to 5 with a fixed $(N, d, \rho) = (1000, 10, 5\%)$.
    Figure \ref{fig:sim_contam_cperp} shows the averaged results over 100 repetitions of the simulation.
    The performance of all methods do not significantly changed as $c_{\perp}/\lambda_1$ increases.
    The robust methods of $g$-DPPCA and NSGGD have better performance than non-robust methods of AG and SGPCA. 
    Importantly, $g$-DPPCA performs better than NSGGD in all cases.
    Meanwhile, SGPCA outperforms among all methods in the case of $c_{\perp}/\lambda_1 = 0.05$, while it becomes the poorest method when $c_{\perp}/\lambda_1 > 0.05$.
    This phenomenon can be explained as follows. 
    Note that the covariance matrix of $X^c$ in \eqref{eq:r2q8-Xc} can be calculated as
    \[
    \cov(X^c) = (1-\rho)\Sigma + \rho\sigma^2I_d + c_{\perp}^2 \cdot \rho(1-\rho) v_{\perp} v_{\perp}^\top.
    \]
    From this expression, one can deduce that for a small $c_{\perp}$ such that 
    \[
    (1-\rho)\lambda_2 + \rho \sigma^2 > c_{\perp}^2 \rho(1-\rho), 
    \]
    the first two PCs of $\cov(X^c)$ are same as those of $\cov(X) = \Sigma$.
    In this case, SGPCA works well since $X^c$ also follows normal distribution.
    However, if $c_{\perp}$ get larger and the above inequality get reversed then SGPCA wrongly estimates PCs as the direction toward $v_{\perp}$ and it fails as in Figure \ref{fig:sim_contam_cperp}.

    For the second additional simulation, we have examined how contamination ratio affects to the methods.
    With same settings as in the previous simulation, we generated contaminated Gaussian data under difference contamination ratios of $\rho \in \{0.025, 0.05, \dots, 0.15\}$ with fixed $(N, d, c_{\perp}) = (2000, 10, 2.5\lambda_1)$.
    The results are shown in Figure \ref{fig:sim_contam_rho}.
    Under $\rho < 0.1$, $g$-DPPCA methods outperform competing methods.
    Interestingly, $g_{wins}$-DPPCA has a better performance than $g_{sph}$-DPPCA in all cases.
    The non-robust methods of AG and SGPCA cannot appropriately carry out PCA in all $\rho$ values.

\subsection{Simulation on non-elliptical distributions}\label{app-sim-nonelliptic}
Here, we have conducted toy simulations to investigate how the proposed methods work on the non-elliptical distributions. We consider two cases. Here, we fix the dimension as $d = 5$ and privacy budget as $(\varepsilon, \delta) = (0.5, 10^{-5})$. Let $u_1 = (1, 0, 0, 0, 0)^\top$ and $u_2 = (0, 1, 0, 0, 0)^\top$ be the first two PCs.
    \begin{enumerate}
        \item (Laplace distribution)
        Independently sample $\ell_1, \dots, \ell_5 \sim \mbox{Lap}(0, 1)$ and generate a random sample as $X = 10\ell_1u_1 + 6\ell_2u_2 + (0, 0, \ell_3, \ell_4, \ell_5)^\top$.
        
        \item (Gamma distribution) Independently sample $r_1, r_2 \sim \mbox{Gamma}(2, \sqrt{2})$ and $\epsilon \sim N_d(0, I_d).$ Then generate a random sample $X$ as $X = r_1u_1 + r_2u_2 + \epsilon$.
    \end{enumerate}

\begin{figure}[]
        \centering
        \includegraphics[width=0.7\linewidth]{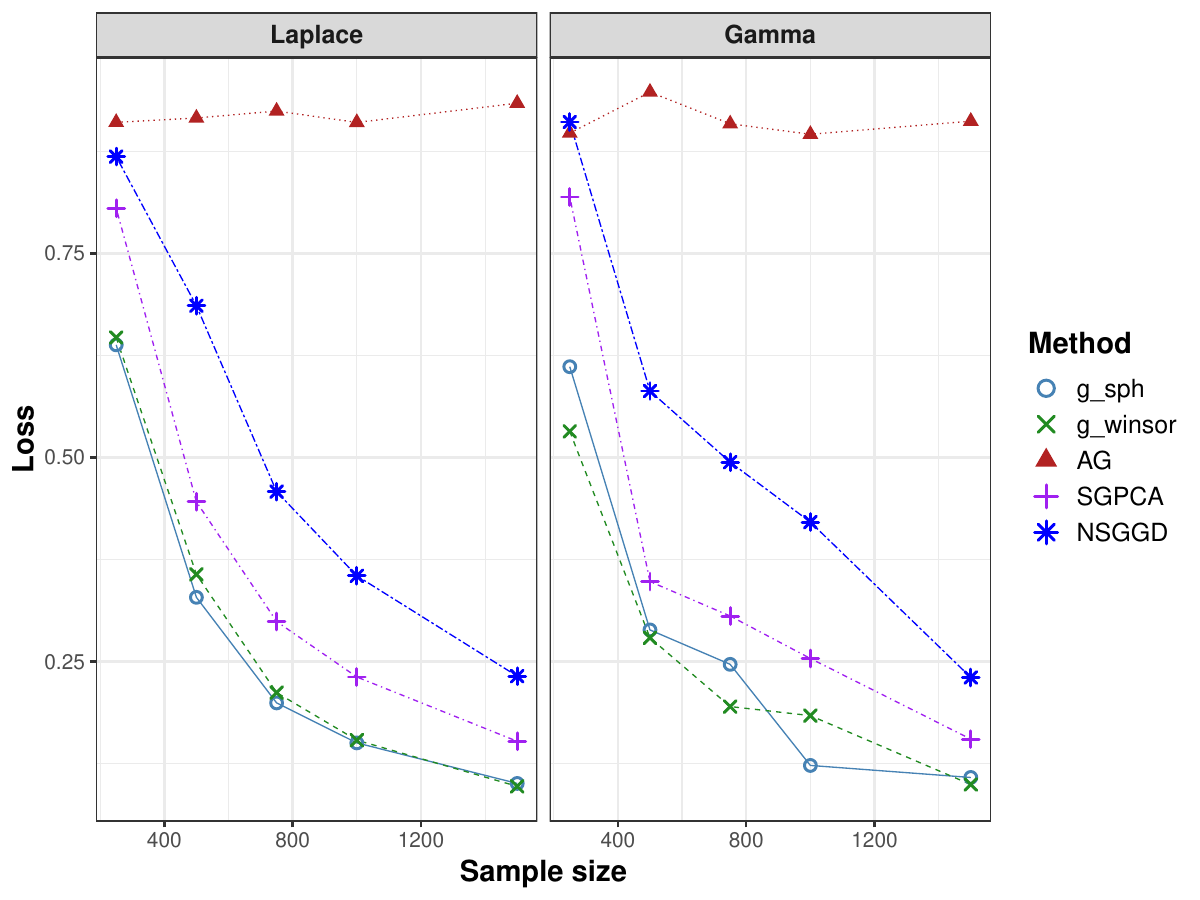}
        \caption{Simulation results for non-elliptically distributed data.}
        \label{fig:nonelliptical}
    \end{figure}

    For each case, we generate $n \in \{250, 500, 750, 1000, 1500\}$ samples and fit the methods to get private PC directions.
    Figure \ref{fig:nonelliptical} shows the results over the 100 repetitions.
    Similar to the elliptically distributed cases, $g_{sph}$-DPPCA and $g_{wins}$-DPPCA outperform the other competing methods.
    So, g-DPPCA can be applied to non-elliptically distributed data and exhibit robustness against various distributional types. 

\section{Technical details for Section \ref{subsec:two-g}} \label{app-3.4}

\renewcommand{\thefigure}{C.\arabic{figure}}
\setcounter{figure}{0}

In this section, we provide proof or verification of conditions appeared in Section \ref{subsec:two-g}.

\subsection{Verification of \texorpdfstring{$\|\Sigma\|_F \log d = \tr (\Sigma) \cdot o(1)$} under spiked covariance} 
\label{subsec-app-c1}

Consider a $m$-spiked covariance matrix $\Sigma = \lambda U_m U_m^\top + \sigma^2 I_d$, where $U_m \in \Oc(d, m)$ and denote a signal to noise ratio as $\ell = \lambda / \sigma^2$.
In this subsection, we verify that the condition $\|\Sigma\|_F \log d = \tr (\Sigma) \cdot o(1)$ in Proposition \ref{prop:errbound-sph} holds when $ \ell \asymp d^{a}  $ for some $ a \in [0, 1) $ under the $m$-spiked covariance matrix model.

Suppose that $ \ell \asymp d^{a}$ for some $a \ge 0$.
After some calculations, we get
\[
    \|\Sigma\|_F = \sqrt{m (\lambda+\sigma^2)^2 + (d-m)\sigma^4}
\]
and $\tr(\Sigma) = m \lambda + d \sigma^2 $. 
Then,
\begin{align*}
\frac{\|\Sigma\|_F \log d}{\tr(\Sigma)} 
= \frac{\log d \sqrt{m (1+\ell)^2 + (d-m)}}{m\ell + d} 
\lesssim \frac{\log d \sqrt{d^{2a} + d}}{d^a + d}.
\end{align*}
Assuming $m = O(1)$, we consider the following cases by the range of $a$:
\begin{enumerate}
    \item Suppose $a \in [0,1/2]$.
    Here, $\sqrt{m\ell^2 + (d-m)} \asymp \sqrt{d^{2a} + d} \asymp  \sqrt{d} $, and 
    \[
    \frac{\log d \sqrt{m \ell^2 + (d-m)}}{m\ell + d} 
    \asymp \frac{\sqrt{d} \log d }{d^{a} + d}
    = \frac{\log d}{d^{a-1/2} + \sqrt{d}}
    = o(1).
    \]
    \item Suppose $a \in (1/2, 1)$. In this case, 
    $\sqrt{m\ell^2 + (d-m)} \asymp \sqrt{d^{2a} + d} \asymp \sqrt{d^{2a}}= d^{a}$ and
    \[
    \frac{\log d \sqrt{m \ell^2 + (d-m)}}{m\ell + d} 
    \asymp \frac{d^{a} \log d }{d^{a} + d}
    = \frac{\log d}{1 + d^{1-a}}
    = o(1).
    \]
    \item Suppose $a \ge 1$. Then 
    \[
    \frac{\log d \sqrt{m \ell^2 + (d-m)}}{m\ell + d} 
    \asymp \frac{  d^a \log d }{d^{a} + d}
    = \frac{\log d}{1 + d^{1-a}}
    \to \infty
    \]
    as $d \to \infty$.
\end{enumerate}
In summary, if $\ell \asymp d^{a}$ for $a \in [0, 1)$, we have $\|\Sigma\|_F \log d = \tr (\Sigma) \cdot o(1)$.

\subsection{Discussion on Condition \texorpdfstring{\eqref{eq-winscond}}{(winscond)} in Proposition \texorpdfstring{\ref{prop:errbound-wins}}{prop:errbound-wins}}
\label{subsec-app-c2}

In this section, we demonstrate Condition \eqref{eq-winscond} is valid for $r \lesssim \sqrt{d\lambda_d}$ under any Gaussian and multivariate $t$ distributions.

Recall that $R^2 \overset{d}{=} \frac{1}{2} (X - \widetilde X)^\top\Sigma^{-1}(X - \widetilde X)$,
where $X \sim \mbox{EC}_d(\mu, \Sigma) $ and $\widetilde X$ is an independent copy of $X$.
Denote $ X \overset{d}{=} \mu + \Sigma^{1/2}Z $ and its independent copy $ \widetilde X \overset{d}{=} \mu + \Sigma^{1/2}\widetilde Z $, where $ Z $ and $ \widetilde Z $ independently follow spherically symmetric distribution. 
Then $ R^2 = \frac{1}{2} (X - \widetilde X)^\top\Sigma^{-1}(X - \widetilde X)   \overset{d}{=} \frac{1}{2} \|Z - \widetilde Z\|_2^2$.
So, $R^2$ is independent of $\Sigma$.

Note that Paley-Zygmund inequality implies that
    \begin{equation*} 
        \Pb(R^2 \ge r^2 / \lambda_d) 
        \ge \left(1-\frac{r^2}{d \lambda_d}\right)^2 \frac{(\Eb R^2)^2}{\Eb R^4}.
    \end{equation*}
    Thus, if $ (\Eb R^2)^2 / \Eb R^4 \asymp 1$ then the condition is satisfied.
    Also, note that
    \[
    \Eb R^2 
    = \frac{1}{2} \Eb \|Z - \widetilde Z\|_2^2  
    = \Eb \|Z\|_2^2 
    \]
    and 
    \begin{align*}
        \Eb R^4
        &= \frac{1}{4} \Eb \left\{ \|Z\|_2^4 + \|\widetilde Z\|_2^4 + 4(Z^\top \widetilde Z)^2 + 2 \|Z\|_2^2\|\widetilde Z\|_2^2 - 4Z^\top \widetilde Z(\|Z\|_2^2 + \|\widetilde Z\|_2^2)\right\}\\
        &= \frac{1}{2} \Eb \|Z\|_2^4 + \frac{1}{2} \Eb \|Z\|_2^2\Eb \|\widetilde Z\|_2^2
        + \Eb (Z^\top \widetilde Z)^2.
    \end{align*}
Based on these observations, we show Gaussian and multivariate $t$ distributions satisfy Condition \ref{eq-winscond}, for $r \lesssim \sqrt{d\lambda_d}$.

\begin{enumerate}
    \item (Gaussian distribution) Let $X \sim N(\mu, \Sigma)$. 
        Then, $ R^2 \overset{d}{=} \|Z\|_2^2 \sim \chi^2_d$, 
        a chi-squared distribution with degree of freedom $d$. 
        In this case, we have 
        $ \Eb R^2 = d $ and $ \Eb R^4 = \var(\chi^2_d) + (\Eb R^2)^2 = 2d + d^2 $.
        Thus, $ (\Eb R^2)^2 / \Eb R^4 \asymp 1$ as $d$ increases.

    \item (Multivariate $t$-distribution) Let $X$ follows a multivariate 
        $t$-distribution with degree of
        freedom $\nu$, i.e., $X$ can be represented as 
        $X \overset{d}{=} \mu + \Sigma^{1/2} W / \sqrt{V/\nu}$, 
        where $W \sim N_d(0, I_d)$ and $V \sim \chi^2_\nu$ are independent. 

        First, consider $ \nu \ge 5 $.
        Then, we have $Z  \overset{d}{=} \frac{W}{\sqrt{V/\nu}}$ and thus
    \[
    \|Z\|_2^2 \overset{d}{=} d \cdot \frac{\chi^2_d/d}{\chi^2_\nu/\nu} 
    \overset{d}{=}  d \cdot F_{d, \nu},
    \]
    where $ F_{d, \nu}$ denotes a $F$-distribution with  $d$ and $\nu$ degrees of freedom. 
    Then we obtain
    \begin{align*}
        \Eb \|Z\|_2^2 &= \frac{\nu}{\nu - 2} \cdot d, \\
    \Eb \|Z\|_2^4 
      &= \frac{\nu^2 }{(\nu-2)(\nu-4)} \cdot d(d+2), \\
    \Eb (Z^\top \widetilde Z)^2 &= 
    \left(\frac{\nu}{\nu-2}\right)^2 \cdot d
    \end{align*}
    Then, it can be easily checked that  $ (\Eb R^2)^2 / \Eb R^4 \asymp 1$.

    Next, we numerically verify that the condition is satisfied for  the $t$-distributions with the degrees of freedom $\nu < 5$. It is enough to investigate the most heavy-tailed case of $\nu = 1$.
    Under the case of $t_1$, we 
    provide numerical evidence that it is enough to choose $ r \lesssim \sqrt{d\lambda_d} $. 
 
    \begin{figure}[ht]
        \centering
        \includegraphics[width=0.6\linewidth]{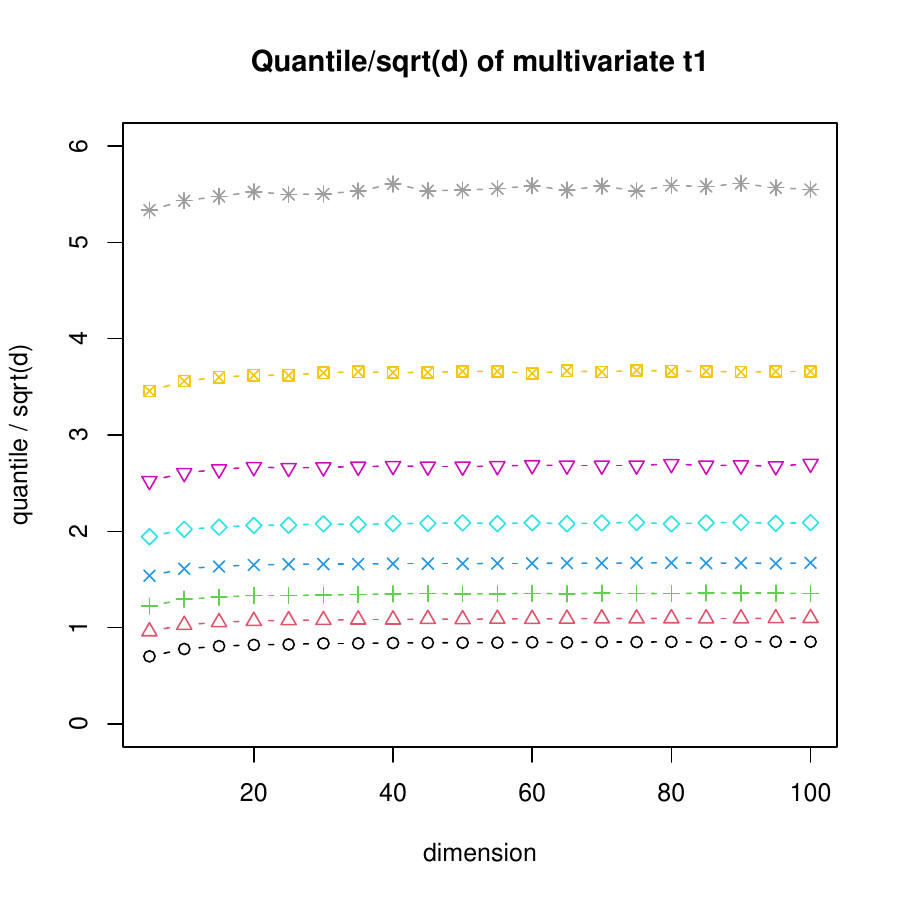}
        \caption{Each line represents the 0.1 to 0.9 quantiles of $R$ as the dimension 
        varies, where $X$ and $\widetilde X$ are sampled from the multivariate $t_1$ distribution.}
        \label{fig:t1_quantiles}
    \end{figure}

    To see this, we first fix $d$ from $\{25, 50, \dots, 200\}$ and generate 100,000 random samples of $\{R_i\}$ where $R_i$ follows $R \overset{d}{=}  \sqrt{(X-\widetilde X)^\top \Sigma^{-1}(X-\widetilde X)/2}$ and $X, \widetilde X \overset{ind}{\sim} t_1(\Sigma)$, or equivalently $ R \overset{d}{=} \|Z-\widetilde Z\|_2/\sqrt{2} $, where $Z, \widetilde Z \overset{ind}{\sim} t_1(I_d)$.
    After that, we obtain $q$-quantile of $\{R_i\}$ for each $q \in \{0.1, 0.2, \dots, 0.9\}$.
    The results are shown in Figure \ref{fig:t1_quantiles}.
    Each line in the figure denotes the $q$-quantiles of $\{R_i\}$ as dimension increases.
    Notably, each line becomes almost constant. 
    This implies that the choice of $r \propto \sqrt{d\lambda_d}$ makes the sense.
    For example, if we take $r = 2\sqrt{d\lambda_d}$, then (3.4) is satisfied since 
    \[
    \Pb(R^2 \ge r^2/\lambda_d)
    = \Pb(R \ge 2\sqrt{d})
    \approx 0.5,
    \]
    where the last approximation comes from the results shown in Figure \ref{fig:t1_quantiles}.

\end{enumerate}

\subsection{Derivation of bounds appeared in Remark \ref{remark_3-1}}
\label{subsec-app-c3}

Here, we derive upper bounds of \eqref{rmk3.1-sph} of $g_{sph}$-DPPCA and SGPCA of \cite{cai2024optimal} appeared in the remark.

Suppose that $ \ell = \lambda / \sigma^2 \asymp d^{a} $ for some $ a \in [0, 1) $.
Note that 
\[
    \mbox{er}(\Sigma) 
    = \frac{\mbox{tr}(\Sigma)}{\lambda_1}
    = \frac{m\lambda + d \sigma^2}{\lambda + \sigma^2}
    = \frac{m\ell + d}{\ell + 1}.
\]
Hence, we have
\[
\mbox{er}(\Sigma) \asymp 
\frac{d^{a} + d}{d^{a} + 1} \asymp d^{1-a}.
\]
From this, the bound \eqref{rmk3.1-sph} becomes 
\[
    O\left(
        \frac{d^{\frac{3}{2} - a}}{\varepsilon n} + \sqrt{\frac{d^{1-a}}{n}}
    \right).
\] 
On the other hand, the bound \eqref{rmk3.1-cai} becomes
\[
    O\left(
    d^{-\frac{a}{2}}
    \left(\frac{d}{\varepsilon n} + \sqrt{\frac{d}{n}}\right) 
    \right)
    =
    O\left(
    \frac{d^{1 - \frac{a}{2}}}{\varepsilon n} + \sqrt{\frac{d^{1-a}}{n}}
    \right).
\]
So, the first term of \eqref{rmk3.1-sph} becomes $ \frac{d^{\frac{3}{2} - a}}{\varepsilon n} $ which is $ d^{\frac{1-a}{2}} $ times larger than $\frac{d^{1 - a}}{\varepsilon n}$, the first term of \eqref{rmk3.1-cai}.
We note that two bounds become same as $a \to 1$.

\section{Implementation details for competing methods}\label{app-C}

\subsection{Analyze Gauss}

\cite{dwork2014analyze} proposed an additive Gaussian mechanism called ``Analyze Gauss'' that adds a Gaussian noise directly to the sample covariance matrix. 
Since the domain of data is assumed to be finite in their work, it does not match our data setting.
To ensure finite sensitivity, we first perform centering the given dataset, and then normalize it by the maximum norm of the centered dataset. 
This enables to use additive Gaussian mechanism, which is described in Algorithm \ref{alg:AG}.

\vspace{10pt}
\begin{algorithm}[H]
\SetAlgoLined
\DontPrintSemicolon
\KwIn{dataset $S = (X_1, \dots, X_n)$; the number of PCs $ m > 0 $; privacy parameters $\varepsilon > 0$, $\delta \in (0, 1)$.}
\KwOut{$(\varepsilon, \delta)$-DP estimate of $\Vc_m(\Sigma)$}
\BlankLine

Center the dataset: $ Z_i \gets X_i - \bar X $ \;

Normalize the dataset: $ Z_i \gets Z_i / \max_i \|Z_i\|_2 $ \;

Compute the sample covariance matrix: 
$\widehat \Sigma \gets (n-1)^{-1} \sum_{i=1}^n Z_i Z_i^\top\;$

Sample Gaussian noise $ \xi \sim N\left(0, \sigma_{\varepsilon, \delta}^2 I_{d(d+1)/2} \right) $, 
where
\[
    \sigma_{\varepsilon, \delta} = \frac{6 \sqrt{2 \ln(1.25/\delta)}}{n \varepsilon}\;
\]

Compute noisy top-$m$ eigenvectors after adding Gaussian noise:
\begin{equation*}
\widetilde V_m \gets \Vc_m\left(\widehat \Sigma + \vecd^{-1}(\xi) \right) \;
\end{equation*}

\Return $\widetilde V_m$\;
\caption{Analyze Gauss \citep{dwork2014analyze}}
\label{alg:AG}
\end{algorithm}
\vspace{10pt}

\begin{cor}
Output $ \widetilde V_m(S) $ of Algorithm \ref{alg:AG} satisfies $ (\varepsilon, \delta) $-DP.
\end{cor}

\begin{proof}
By Proposition \ref{prop:gaussian-mech}, it is enough to show that the sensitivity of 
$ \widehat \Sigma $ appeared in line 3 of the algorithm is bounded by $ 6 / n $.
Denote $ R = \max_i \|X_i - \bar X\|_2 $.
Then $ \widehat \Sigma(S) $ can be written as 
\[
\widehat \Sigma(S) = \frac{1}{R^2(n-1)} \sum_{i=1}^{n} (X_i - \bar X) (X_i - \bar X)^\top.
\]
By using an $U$-statistic representation, we have
\[ 
    \widehat{\Sigma}(S) = \frac{1}{R^2} \cdot \frac{1}{n(n-1)} \sum_{i < j} (X_i - X_j)(X_i - X_j)^\top.
\]
Consider a neighboring dataset $S'=(X_1, \dots, X_n')$ of $ S $. 
Then, we have
\begin{align*}
\|\widehat{\Sigma}(S) - \widehat{\Sigma}(S')\|_F
&\le \frac{1}{R^2n(n-1)} 
\bigg\Vert\sum_{i=1}^{n-1} (X_n - X_i)(X_n - X_i)^\top - (X_n' - X_i)(X_n' - X_i)^\top \bigg\Vert_F \\
&\le \frac{1}{R^2n(n-1)} \cdot (n-1) \cdot 6R^2  
\end{align*}
This shows that $\Delta_F(\widehat{\Sigma}) \le 6 / n$ and completes the proof.
\end{proof}

\subsection{SGPCA}
Here, we describe the algorithm and parameter setting we used for SGPCA. 
The following algorithm follows from Algorithm 1 of \cite{cai2024optimal}.

\vspace{10pt}
\begin{algorithm}[H]
\SetAlgoLined
\DontPrintSemicolon
\KwIn{dataset $S = (X_1, \dots, X_n)$; eigenvectors sensitivity $\Delta$; the number of PCs $ m > 0 $; privacy parameters $\varepsilon > 0$, $\delta \in (0, 1)$.}
\KwOut{(approximate) $(\varepsilon, \delta)$-DP estimate of $\Vc_m(\Sigma)$}
\BlankLine

Get pairwise dataset: $ Z_i \gets (X_{n/2 + i} - X_i) / \sqrt{2} $ for $ i = 1, \dots, n/2 $\;

Compute the sample covariance matrix and top-$m$ eigenvectors:
\[
\widehat \Sigma \gets \frac{1}{n} \sum_{i=1}^n Z_i Z_i^\top \quad \text{ and } \quad
\widehat V_m \gets \Vc_m(\widehat \Sigma)\;
\]

Compute noisy top-$m$ eigenvectors after adding Gaussian noise:
\begin{equation*}
\widetilde V_m \gets \Vc_m\left(\widehat{V}_m\widehat{V}_m^\top + E \right)
\end{equation*}
where 
\[
E_{ij} = E_{ji} \overset{i.i.d.}{\sim} N\left(0, \frac{2\Delta^2}{\varepsilon^2} 
\log \frac{1.25}{\delta} \right),\quad \forall 1 \le i \le j \le d \;
\]

\Return $\widetilde V_m$\;
\caption{Private Spiked covariance Gaussian PCA \citep{cai2024optimal}}
\label{alg-sgpca}
\end{algorithm}
\vspace{10pt}

For our implementation, we follow the parameter settings described in Section 6.1 of \cite{cai2024optimal}. 
Although the rank and eigenvalues of $\Sigma$ are unknown, we assume they are known a priori in our 
simulation setting for SGPCA. 
Specifically, we set the sensitivity of eigenvectors, $\Delta$, to
\[
\Delta = 4 \left(\frac{\lambda_d}{\lambda_1}  + \sqrt{\frac{\lambda_d}{\lambda_1}}\right) \frac{\sqrt{d(r+\log n)}}{n},
\]
and the rank $r$ to $r = m$. 
We note that the sensitivity bound holds with a high probability, implying that SGPCA does not guarantee the exact $(\varepsilon, \delta)$-DP.

\subsection{NSGGD} 
Here, we present the geodesic descent type algorithms proposed in \cite{maunu2022stochastic}.
Detailed algorithm is described in Algorithm \ref{alg:maunu-nsggd}.
The algorithm utilizes the idea of private stochastic descents.
In our numerical studies, we set $ T = n^2 $ and $ B = \max (n \sqrt{\frac{\varepsilon}{8T}}, 1) $.
We note that convergence speed of Algorithm \ref{alg:maunu-nsggd} heavily depends on the learning rates $\eta_k$'s.
We use $ \eta_k = 1/n^2 $ since this works well in our numerical study among some candidates.

\begin{algorithm}[t]
\SetAlgoLined
\DontPrintSemicolon
\KwIn{dataset $S = (X_1, \dots, X_n)$; 
    the number of PCs $m > 0$; privacy parameters $\varepsilon, \delta \in (0, 1)$; 
    batch size $ B $; number of iteration $ T $; learning rates $ \{\eta_k\}_{k=1}^T $
}
\KwOut{$(\varepsilon, \delta)$-DP estimate of $\Vc_m(\Sigma)$}
\BlankLine
Set $ S \gets (z_1, \dots, z_{n/2}) $, where $ z_i = (X_{n+i} - X_i) / \|X_{n+i} - X_i\|_2$, 
and $ n \gets n/2 $\;

Set $ \varepsilon' = \varepsilon/2 $ and $ \delta' = \delta/2 $ \;

Set $ \sigma \gets B\sqrt{2 T \log (1/\delta')} / (n^2\varepsilon')  $\;

Set the initial point $ V_0 \in \Pc(d, m) $ by the output of $ (\varepsilon', \delta') $-DP Analyze Gauss mechanism \;

\For{$ k = 0 $ to $ T-1 $}{
    \BlankLine
    Randomly sample batch $ S_k \subset S $ with batch size $ |S_k| = B $ with replacement\;
    Set projection matrix $ Q_{k} \gets I - V_{k}V_{k}^\top$\;
    Get batch gradient at $ V_k $ as $ \nabla F(V_{k}; S_k) \gets \frac{1}{B} \sum_{x \in S_k} \frac{Q_k xx^\top V_k}{\|Q_k x\|_2}  $\;
    Set $ E_k \in \mathbb{R}^{d \times m} $ whose entries are i.i.d. samples from  $ N(0, \sigma^2) $ \;
    Update $ V_{k+1} \gets P_{\mathcal{O}(d, m)}\left(V_{k} - \eta_k(\nabla F(V_{k}; S_k) + E_k)\right) $; \;
    \quad Here, 
    $ P_{\mathcal{O}(d, m)}(A) := \argmin_{V \in \Oc(d, m)} \|V-A\|^2 = UW^\top $, whenever 
    $ A = U\Sigma W^\top $ is the SVD of $ A $.\;
}

\Return $V_T$\;

\caption{Noisy Stochastic Geodesic Gradient Descent}
\label{alg:maunu-nsggd}
\end{algorithm}

\end{document}